\documentclass[letterpaper,onecolumn,12pt]{article}
\usepackage{usenix2019_v3}

\usepackage{tikz}
\usepackage{amsmath}
\usepackage{amssymb}
\usepackage{amsthm}
\usepackage{graphicx}
\usepackage{hyperref}
\usepackage{changepage}
\usepackage{parskip}
\graphicspath{ {./graphs/} }

\usepackage{math-macros}
\usepackage{braket}
\usepackage{appendix}

\begin{document}

\date{}

\begin{centering}\title{\begin{centering}\Large \bf Quantum Coloring Games on Complete Graphs\end{centering}}\end{centering}

\author{
{\rm Will Bolduc}\\
The University of Texas at Austin
\and
{\rm Michael Jennings}\\
The University of Texas at Austin
} 
\maketitle

\begin{abstract}

We study the nonlocal game where two players are both given an input ($x$ and $y$, respectively), from a set of size $n$ and, without communicating, must give outputs $a$ and $b$, respectively, from a set of size $m < n$, winning when $(x = y) \leftrightarrow (a = b)$. This can be considered a graph coloring game on the complete graph with $n$ vertices. In the simplest case, the players get random equal inputs with probability $\frac{1}{2}$ and random unequal inputs otherwise.

We solve this case of the game, and many others, for both classical and quantum (i.e., where the players may share finite-dimensional entangled quantum states) strategy sets, proving the value (the probability of winning) of the optimal strategies and providing explicit strategies that attain this optimum for all $n$ and $m$. 

Our optimal strategies are \textit{synchronous}: i.e., the two players always give equal outputs when given equal inputs. These strategies remain optimal when the probability of receiving equal inputs is at least $\frac{2}{m + 2}$. 
Across this wide range of cases,
we find that quantum strategies have an advantage over classical strategies exactly when $m$ does not divide $n$; this advantage is $O\lr{\frac{m}{n^2}}$.

\end{abstract}

\section{Background}

\subsection{Nonlocal Games}
A nonlocal game is a cooperative game, traditionally with two players named Alice and Bob and a single referee (also sometimes referred to as a verifier). 
The referee chooses inputs for each of Alice and Bob from a finite set of inputs via some fixed probability distribution. Alice and Bob, once they receive their inputs, may not communicate in any way, and must give outputs to the referee. The referee decides whether or not the game has been won according to some predicate function of the inputs and outputs. 

The \textit{value} of a nonlocal game on a class of strategies is defined as the supremum of the winning probabilities of all such strategies. The \textit{quantum value} is the value on the class of quantum strategies (where Alice and Bob can share any finite amount of entanglement), and the \textit{classical value} is the value on the class of classical strategies. The \textit{quantum advantage} of a game can then be defined as the difference between these two quantities.

Interest in the quantum values of nonlocal games was popularized by Cleve, H\o yer, Toner, and Watrous \cite{calons}. To use a famous example, in the CHSH game \cite{pettlhvt}, Alice and Bob each receive a single bit and must each output a single bit, winning if the parity (i.e., XOR) of their outputs equals the logical AND of their inputs. The classical value of this game is $\frac{3}{4} = 0.75$, whereas the quantum value is $\cos^2\lr{\frac{\pi}{8}} \approx 0.85$, so the game has positive quantum advantage. These games have been proposed as experiments to refute ``local hidden variable'' theories of quantum mechanics \cite{oteprp}, which would preclude any quantum advantage. Some such experiments have been successfully run \cite{bellexperiment}.

Building upon this work was the study of XOR games. An XOR game is defined as one where Alice and Bob may each respond with only a single bit, and the referee considers only the inputs and the parity (i.e., XOR) of the outputs when deciding whether the players won. It can be quickly observed that the CHSH game is an XOR game, as indeed is the $m = 2$ case of the game we study. XOR games have received a substantial amount of study, from investigating their quantum values to bounding the entanglement needed to play optimally, and there is notably a method to compute the optimal solution for an XOR game via a semidefinite program \cite{calons, elaubotevomxg}.

One interesting variation on the study of nonlocal games, and one that much of our paper will discuss, is analyzing the \textit{synchronous value} of a nonlocal game \cite{svog}.  A \textit{synchronous strategy} is a strategy that guarantees that, if Alice and Bob are given the same input, then they will respond with the same output. Note that if we repeatedly give Alice and Bob the same input across multiple ``rounds'', we do not require that they give the same output for each ``round'', only that in each ``round'' Alice's output matches Bob's output. The \textit{synchronous value} of a game for a class of strategies is defined as the supremum of the success probabilities of all \textit{synchronous strategies} in that class. 

Games that require the players to answer (classically impossible) questions about graphs have garnered significant recent research interest. One example of this is the study of the \textit{chromatic number} and \textit{quantum chromatic number} of a graph \cite{qcng}. Both can be defined relative to a nonlocal game where Alice and Bob seek to convince a referee that they have a valid $m$-coloring of a graph $G=(V,E)$. The referee asks each of Alice and Bob what color a specific vertex ($v_a$ and $v_b$, respectively) is. If given the same vertex, they must respond with the same color, and if given adjacent vertices, they must respond with different colors. The \textit{chromatic number} of a graph $G$ is defined as the minimum number of colors for which this graph coloring game has classical value $1$. Likewise, the \textit{quantum chromatic number} of $G$ is the minimum number of colors where the graph coloring game has quantum value $1$. Notably, not every graph's quantum chromatic number equals its classical chromatic number: Cameron et al. give a graph with classical chromatic number $5$ but quantum chromatic number $4$ \cite{qcng}. The graph coloring game on specifically complete graphs has received some special attention, which we will discuss in \S\ref{sec:prevwork}.

Graph theory has also been used as a tool to investigate values of nonlocal games. In particular, Chailloux et al. construct a ``game graph'' encoding the verifier's predicate \cite{gtbotevonlg}, then bound both the classical and quantum values of a game by an efficiently computable quantity of this graph known as the Lov\'asz theta number \cite{gtbotevonlg, otscoag}.

Nonlocal games, and the quantum advantages that are achievable in some of them, have also shown some relevance to networking. Previous work has investigated how the superior correlation offered by shared entanglement could be used to improve coordination and load balancing in distributed or networked systems \cite{ftlcfnswqnlg, eicids}.

\subsection{The $m$-Coloring Game on an $n$-Node Complete Graph}

In this paper, we will focus on the $m$-coloring game on the complete graph $K_n$. More precisely, Alice and Bob are given inputs $x$ and $y$, respectively, from a set of size $n$ (which we will represent as $[n] := \{k \in \mathbb{N}_0 \mid k < n\}$) and must give outputs $a$ and $b$, respectively, from a set of size $m$ (which we will represent as $[m]$), winning when $(x = y) \leftrightarrow (a = b)$.

To preserve symmetry, we will consider only input distributions $\pi$ over $[n] \times [n]$ that assign equal probability to every pair of inputs where $x\neq y$ and equal probability to every pair of inputs where $x=y$: i.e., letting $p := \mathbb{P}[x = y]$, we have $\pi(x, x) = \frac{p}{n}$ for each $x \in [n]$ and $\pi(x, y) = \frac{1 - p}{n(n - 1)}$ (or 0 when $n=1$) whenever $x \neq y$.

When $m \ge n$, Alice and Bob may always simply output their respective inputs, in which case the win condition becomes $(x = y) \leftrightarrow (x = y)$, which is trivially always true and hence this strategy is clearly optimal. Also, when $m = 1$, Alice and Bob both must output $0$ and thus have no choice in strategy. Hence, we focus on the interesting case where $2 \le m < n$. 

Similarly, when $p = 1$, the players always win if Alice and Bob both output $0$ regardless of their inputs, and when $p = 0$ they always win if Alice always outputs $0$ and Bob always outputs $1$ (which is valid for all $m \ge 2$), so we focus on the interesting case where $p \in (0, 1)$.

\subsection{Applications in Distributed Computer Systems}
This game can also be interpreted in a load-balancing scenario: imagine a networked system with $m$ machines and $n$ users. Users can submit jobs (queries that specify a user $x$ and $y$ respectively, which each correspond to one of the $n$ vertices of $K_n$) to two load balancers (Alice and Bob) far enough apart that classical communication between them is not viable. The load balancers must return a machine ($a$ and $b$ respectively, each corresponding to one of the $m$ colors) on which the job should run. We prefer for jobs for the same user to run on the same machine, so if both jobs are for the same user (i.e., $x = y$), then we will consider it a success if $a=b$. Jobs for different users should be spread across different machines, so if $x \neq y$, a success would require that $a \neq b$.

This framing also helps to justify the interest in $p$, the probability that Alice and Bob receive the same input. Increasing $p$ can be seen as increasing the relative weight in the objective of equal-input scenarios, effectively increasing the penalty for routing one user's jobs to two different machines relative to the penalty for placing two unrelated jobs on the same machine. Similarly, synchronous strategies for this game (i.e., strategies that guarantee that $x=y\implies a=b$) can be thought of as strategies for the scenario where a given user's jobs submitted in the same round absolutely must run on the same machine.

\subsection{Previous Work}\label{sec:prevwork}
There are a few notable past results that will guide our work here. The first comes from Cameron et al., which states that the quantum chromatic number of a complete graph $K_n$ is $n$: that is, it is impossible to always convince a verifier (i.e., win our game) of a coloring with $m$ colors for any $m < n$. Thus, they show that the game we study has no perfect quantum strategy. We prove stronger results, bounding the value more tightly and finding the exact value in many cases.

Secondly, Cui et al., in their analysis of the Bipartite Perfect Matching game, establish that there is a quantum advantage for the $2$-coloring game played on a uniform input distribution on $K_3$, but that there is no advantage for $K_n$ when $n \geq 4$ \cite{qpm}. This solves a special case of our game where $m = 2$ and $p=\frac{1}{n}$. 

Finally, Helton et al. bound the optimal synchronous value of the coloring game (on arbitrary graphs, not just complete ones) in terms of 
the \textit{graph correlation function} \cite{ncotsoqcvg}, which they compute explicitly for (most) complete graphs \cite{svog}. 
We use these theorems to prove that our quantum strategy is optimal in the synchronous case.

\subsection{Main Results} Our results all concern input distributions where the players receive uniformly random equal inputs with probability $p$ and uniformly random unequal inputs with probability $1 - p$.

In \S\ref{sub:ourclassicalstrategy}, we prove that the synchronous classical value of the game we study is $$p + (1 - p)\lr{1 - \frac{(\left\lfloor\frac{n}{m}\right\rfloor - 1)n + (\left\lfloor\frac{n}{m}\right\rfloor + 1)(n \bmod m)}{n(n - 1)}}.$$
In \S\ref{sub:classicalvalue}, 
we prove that this value is the classical value of the game whenever $p \ge \frac{2}{m + 2}$. 

In \S\ref{sub:ourquantumstrategy} we present a synchronous quantum strategy for this game with value $$p + (1-p)\lr{1-\frac{n-m}{m(n-1)}},$$ which is notably the same expression as the classical value when $n \bmod m = 0$. We prove in \S\ref{sub:quantumsynchronousoptimality} that this strategy is optimal across all synchronous quantum strategies. We then prove in \S\ref{sub:quantumvalue} that the same strategy is optimal across all quantum strategies when $p \geq \frac{2n-m}{(m+2)n-2m}$.

Letting $r$ be the remainder of $n$ divided by $m$, we then show in  \S\ref{sub:quantumadvantage} 
that this optimal strategy achieves an advantage (in terms of difference in value) of $$(1 - p)\lr{\frac{r(m - r)}{mn(n - 1)}}$$
over the best classical synchronous strategies, as well as over the best classical strategies when $p \geq \frac{2}{m+2}$.
Notably, we show that this advantage is $0$ if and only if $m$ divides $n$.

\section{Preliminaries}
We can begin by establishing some helpful results and definitions.

\subsection{Value}
The \textit{value} $\omega$ of a strategy is simply defined as the probability that the players win the game when using that strategy. 
If we define $c_{xy}$ as the probability that $a = b$ given the input pair $(x, y)$, the value can  be written as $$\sum_{x = 0}^{n - 1}\frac{p}{n}c_{xx} + \sum_{x = 0}^{n - 1}\sum_{y \neq x}\frac{1 - p}{n(n - 1)}(1 - c_{xy}) = \frac{p}{n}\sum_{x = 0}^{n - 1}c_{xx} + \frac{1 - p}{n(n - 1)}\sum_{x = 0}^{n - 1}\sum_{y \neq x}(1 - c_{xy}).$$

We can insert a term to counteract the skipped index in the sums over $y$, giving
$$\frac{p}{n}\sum_{x = 0}^{n - 1}c_{xx} - \frac{1 - p}{n(n - 1)}\sum_{x = 0}^{n - 1} (1 - c_{xx}) + \frac{1 - p}{n(n - 1)}\sum_{x = 0}^{n - 1}\sum_{y = 0}^{n - 1}(1 - c_{xy}).$$

Rewriting this expression as a polynomial in the variables $\{c_{xy} \}$ gives
$$\frac{(1 - p)n}{n - 1} - \frac{1 - p}{n - 1} + \lr{\frac{p}{n} + \frac{1 - p}{n(n - 1)}}\sum_{x = 0}^{n - 1} c_{xx} - \frac{1 - p}{n(n - 1)}\sum_{x = 0}^{n - 1}\sum_{y = 0}^{n - 1}c_{xy}.$$

The constant term in this polynomial can be simplified to $1 - p$.

It is then convenient to define $$\xi := \sum_{x = 0}^{n - 1}c_{xx}$$ and $$\mu := \sum_{x = 0}^{n - 1} \sum_{y = 0}^{n - 1}c_{xy}.$$
Since $0 \le c_{xy} \le 1$ for all $(x, y) \in [n]^2$, $0 \le \xi \le n$ and $0 \le \mu \le n^2$. 

Also, $\mu - \xi = \sum_{x = 0}^{n - 1}\sum_{y = 0}^{n - 1}c_{xy} - \sum_{x = 0}^{n - 1}c_{xx} = \sum_{x = 0}^{n - 1}\sum_{y \neq x}c_{xy} \ge 0$, so $\xi \le \mu$.

We can then write the value in terms of these abbreviations: $$\omega = 1 - p + \lr{\frac{p}{n} + \frac{1 - p}{n(n - 1)}}\xi - \frac{1 - p}{n(n - 1)}\mu.$$
We can ignore the constant term during optimization since it does not depend on the strategy. Furthermore,
$1 - p > 0$ whenever $p < 1$, and $n(n - 1) > 0$ for all $n \ge 2$, so we can multiply the rest by $\frac{n(n - 1)}{1 - p}$  without changing signs, giving the new objective 
$$\nu := \lr{\frac{p(n - 1)}{1 - p} + 1}\xi - \mu.$$ 

We can then convert from $\nu$ to the value $\omega$ via $$\omega = 1 - p + \frac{1 - p}{n(n - 1)}\nu.$$
We will now let $s := \frac{p(n - 1)}{1 - p} + 1$ denote the coefficient of $\xi$ in $\nu$; then $$s - ps = s(1 - p) = p(n - 1) + 1 - p = pn - p - p + 1 = p(n - 2) + 1,$$ so $s - 1 = p(n + s - 2)$ and thus $p = \frac{s - 1}{n + s - 2}$.

\begin{mylemma}~\label{lem:smonotone}
    $s$ is monotonically increasing with respect to $p$ on the domain $p \in (0, 1)$.
\end{mylemma}
\begin{proof}
    Since $n > 1$, the numerator $p(n - 1)$ in $\frac{p(n - 1)}{1 - p}$, the non-constant term of $s$, is monotonically increasing in $p$ and positive whenever $p > 0$ and thus all $p \in (0, 1)$.
    
    Also, the denominator $1 - p$ is positive on $p \in (0, 1)$ and monotonically decreasing in $p$.

    Thus, the fraction $\frac{p(n - 1)}{1 - p}$ is positive and increasing in magnitude on $p \in (0, 1)$, so $\frac{p(n - 1)}{1 - p}$, and thus $s = \frac{p(n - 1)}{1 - p} + 1$, are increasing on $p \in (0, 1)$, as desired.
\end{proof}

The coefficient of $\xi$ is positive and the coefficient of $\mu$ is negative for all $p \in (0, 1)$, so any optimal strategy minimizes $\mu$ for its value of $\xi$ and conversely maximizes $\xi$ for its value of $\mu$. 

We sometimes require that Alice and Bob play \textit{synchronous} strategies: i.e., strategies where $\P{a = b \mid x = y} = 1$, which fixes $\xi = \sum_{x = 0}^{n - 1} c_{xx} = \sum_{x = 0}^{n - 1} 1 = n$. In this case, the goal is simply to minimize $\mu$ subject to the synchronous constraint.

\begin{mytheorem}\label{thm:synchoptimalatpmeansoptimalabovep}
    If a synchronous strategy is optimal for some $p^* \in [0, 1]$, it is optimal for all $p \in [p^*, 1]$ as well.
\end{mytheorem}

\begin{proof}
    Since $\xi$ and $\mu$ do not depend on $p$, any strategy's value is linear in $p$.
    
    Any synchronous strategy has value $1$ (and hence is optimal) at $p = 1$.

    Hence, if some synchronous strategy $S$ is optimal at $p = p^*$, any other strategy $S'$ will have nonstrictly lower value at $p = p^*$ and $p = 1$, and thus the value of $S'$ cannot cross above that of $S$ at any $p \in [p^*, 1]$. 
\end{proof}

\subsection{Mixed Strategies} 
We allow the players to use mixed (i.e., random) strategies, and since Alice and Bob can communicate before receiving their inputs, they may have access to shared random values, so their choices of strategies may not be independent. 

Examples of mixed strategies in our game include classical random functions, randomly chosen measurement operators, and shared random classical values or mixed quantum states.

Formally, we can describe mixed strategies as distributions over the space of pure strategies.

However, the value of such a strategy is then a convex combination (weighted by probability) of values of pure strategies and hence cannot exceed the supremum of those values. 

Thus, when searching for optima, we need only consider pure strategies. 
\section{Classical Strategies}

When giving their respective outputs, each player knows only their own input, and has no information about the other player's input or output (beyond the prior of their strategy).

A pure classical strategy for this game is then given by a pair $(a, b) \in ([n] \to [m])^2$ of deterministic functions  specifying the players' outputs in response to each input.

\subsection{Values of Classical Strategies}
The value of such a strategy pair can be computed as $$\omega(a, b) = \mathbb{P}[a(x) = b(y) \leftrightarrow x = y \mid (x, y) \overset{\pi}{\leftarrow} [n]^2].$$ 

Alternatively, $c_{xy} := \mathbb{P}[a = b \mid (x, y)] = 1$ if $a(x) = b(y)$ and $c_{xy} = 0$ otherwise, so $$\xi := \sum_{x = 0}^{n - 1}c_{xx} = |\{x \in [n] \mid a(x) = b(x)\}|$$ and
$$\mu := \sum_{x = 0}^{n - 1}\sum_{y = 0}^{n - 1}c_{xy} = |\{(x, y) \in [n]^2 \mid a(x) = b(y)\}|.$$

Partitioning the former set according to output value gives $\xi = \sum_{k = 0}^{m - 1}\xi_k$, where $$\xi_k := |\{x \in [n] \mid a(x) = b(x) = k\}|.$$
Partitioning the latter set according to output value gives $\mu = \sum_{k = 0}^{m - 1}\mu_k$, where $$\mu_k := |\{(x, y) \in [n]^2 \mid a(x) = b(y) = k\}| = |\{x \in [n] \mid a(x) = k\}| \cdot |\{y \in [n] \mid b(y) = k\}|.$$ 
Defining $a_k := |\{x \in [n] \mid a(x) = k\}|$ and $b_k := |\{y \in [n] \mid b(y) = k\}|$ gives $\mu = \sum_{k = 0}^{m - 1}a_kb_k$.

\begin{mylemma}~\label{lem:mukleakandbk}
    For any output $k$, we must have $\xi_k \le a_k$ and $\xi_k \le b_k$.
\end{mylemma}
\begin{proof}
    $\{x \in [n] \mid a(x) = b(x) = k\} \subseteq \{x \in [n] \mid a(x) = k\}$, so $\xi_k \le a_k$.

    Likewise, $\{x \in [n] \mid a(x) = b(x) = k\} \subseteq \{x \in [n] \mid b(x) = k\}$, so $\xi_k \le b_k$.
\end{proof}

\begin{mylemma}\label{lem:mukatleastxikssquared} 
    Any valid classical strategy has $\mu \ge \sum_{k = 0}^{m - 1}\xi_k^2$.
\end{mylemma}
\begin{proof}
    By Lemma \ref{lem:mukleakandbk}, $\xi_k^2 = \xi_k\xi_k \le a_kb_k = \mu_k$ for each $k$, so $\mu = \sum_{k = 0}^{m - 1} \mu_k \ge \sum_{k = 0}^{m - 1} \xi_k^2$.
\end{proof}

\begin{mylemma}\label{lem:xikssquaredminimizer}
    For fixed $\xi = \sum_{k = 0}^{m - 1}\xi_k$, the quantity $\sum_{k = 0}^{m - 1}\xi_k^2$ is minimized when exactly $\xi \bmod m$ of the $\xi_k$ equal $\left\lfloor\frac{\xi}{m}\right\rfloor + 1$, and the rest equal $\left\lfloor\frac{\xi}{m}\right\rfloor$.
    \end{mylemma}
\begin{proof}
    If $\xi_h - \xi_l > 1$ for some $(h, l) \in [m]^2$, we could take any $x$ for which $a(x) = b(x) = h$ (which must exist since $\xi_h > 1 + \xi_l \ge 1$) and set $a(x) = b(x) = l$ instead. 
    
    Then $\xi_h$ decreases by $1$, 
    $\xi_l$ increases by $1$, 
    and no other $\xi_k$ changes, which adds $$(\xi_h - 1)^2 - \xi_h^2 + (\xi_l + 1)^2 - \xi_l^2 = -2\xi_h + 1 + 2\xi_l + 1 = 2(\xi_l + 1 - \xi_h) < 2 \cdot 0 = 0$$ to $\sum_{k = 0}^{m - 1}\xi_k^2$: strategies whose values of $\xi_k$ differ by more than $1$ cannot attain the minimum.
    
    Any strategy attaining the minimum then has $\xi_k \in \{q_\xi, q_\xi + 1\}$ for all $k \in [m]$ for some $q_\xi$.
    
    We can further assume without loss of generality that $\xi_k = q_\xi$ for some $k \in [m]$, since if all $\xi_k = q_\xi + 1$ we could just choose $q_\xi$ to be $q_\xi + 1$. Then letting $r_\xi := |\{k \in [m] \mid \xi_k = q_\xi + 1\}|$, $$
        \xi = \sum_{k = 0}^{m - 1} \xi_k
        = (m - r_\xi)q_\xi + r_\xi(q_\xi + 1)
         = mq_\xi - r_\xi q_\xi + r_\xi q_\xi + r_\xi = mq_\xi + r_\xi,
    $$
    and since $r_\xi$ is the cardinality of a proper subset of $[m]$, $r_\xi \in [0, m)$.
    
    Therefore, by the Division Algorithm, $r_\xi = \xi \bmod m$ and $q_\xi = \left\lfloor\frac{\xi}{m}\right\rfloor$, as desired.
\end{proof}

\begin{mylemma}\label{lem:xikssquaredminimum} 
    The minimum value of $\sum_{k = 0}^{m - 1}\xi_k^2$ for any fixed $\xi = \sum_{k = 0}^{m - 1}\xi_k$ is exactly $$\left\lfloor\frac{\xi}{m}\right\rfloor^2(m - (\xi \bmod m)) + \lr{\left\lfloor\frac{\xi}{m}\right\rfloor + 1}^2(\xi \bmod m).$$
\end{mylemma}
\begin{proof}
    This is the value of $\sum_{k = 0}^{m - 1}\xi_k^2$ attained by the minimizer, as proven in Lemma \ref{lem:xikssquaredminimizer}.
\end{proof}

\subsection{Synchronous Classical Strategies}~\label{sub:ourclassicalstrategy}
In order to satisfy the definition that Alice and Bob always give equal outputs when given equal inputs, synchronous strategies must have $a(x) = b(x)$ for all $x \in [n]$ (i.e., $a = b$), so $$\{x \in [n] \mid a(x) = b(x) = k\} = \{x \in [n] \mid a(x) = k\} = \{x \in [n] \mid b(x) = k\}$$ and thus $\xi_k = a_k = b_k$, so $\mu = \sum_{k = 0}^{m - 1} a_kb_k = \sum_{k = 0}^{m - 1} \xi_k \xi_k = \sum_{k = 0}^{m - 1} \xi_k^2$ exactly. 

Also, $\xi = \sum_{k = 0}^{m - 1} \xi_k = \sum_{k = 0}^{m - 1} a_k = n$, confirming that such strategies are indeed synchronous.

We now define $r := n \bmod m$ and $q := \left\lfloor\frac{n}{m}\right\rfloor$, so that $n = qm + r$ and $r \in [0, m)$. 
\begin{mytheorem}~\label{thm:ourclassicalstrategymu}
    The optimal classical synchronous strategy has $\mu = q^2(m - r) + (q + 1)^2r$.
\end{mytheorem}
\begin{proof}
    For synchronous strategies, $\mu = \sum_{k = 0}^{m - 1} \xi_k^2$ exactly and $\xi = n$, so by Lemma \ref{lem:xikssquaredminimum}, the minimum value of $\mu$ across all synchronous strategies is exactly $q^2(m - r) + (q + 1)^2r$. \end{proof}    

\begin{myremark}
    One explicit strategy that attains this $\mu$ has $a(x) = b(x) = x \bmod m$ for all $x \in [n]$.
\end{myremark}

\begin{myremark}
    The optimal synchronous strategy distributes inputs across outputs as uniformly as possible, which parallels the saturation condition for the Cauchy-Schwarz inequality invoked in bounding the quantum optima, except that indivisibility may prevent the numbers of inputs allocated to each output from being exactly equal in the classical case.
\end{myremark}

\begin{mytheorem}~\label{thm:ourclassicalstrategyvalue}
    The classical synchronous value of this game is $$p + (1 - p)\lr{1 - \frac{(q - 1)n + (q + 1)r}{n(n - 1)}}.$$
\end{mytheorem}
\begin{proof}
Recall that we can write the strategy's value in terms of $\xi$ and $\mu$: $$     \omega = 1 - p + \lr{\frac{p}{n} + \frac{1 - p}{n(n - 1)}}\xi - \frac{1 - p}{n(n - 1)}\mu.$$
    By Theorem \ref{thm:ourclassicalstrategymu}, the optimal classical synchronous strategy has $\mu = q^2(m - r) + (q + 1)^2r$, so $$\begin{aligned}
    \omega 
    & = 1 - p + \lr{\frac{p}{n} + \frac{1 - p}{n(n - 1)}}n - \frac{1 - p}{n(n - 1)}(q^2(m - r) + (q + 1)^2r) \\
    & = 1 - p + p + \frac{1 - p}{n - 1} - \frac{1 - p}{n(n - 1)}(q^2(m - r) + (q + 1)^2r) \\
    & = p + (1 - p)\lr{1 + \frac{n - (q^2(m - r) + (q + 1)^2r)}{n(n - 1)}}.
    \end{aligned}$$
    Since $n = mq + r$, we can simplify the term $q^2(m - r) + (q + 1)^2r$ to
    $$q^2m - q^2r + (q^2 + 2q + 1)r = q^2m + 2qr + r = q(mq + r) + qr + r = qn + (q + 1)r.$$
    Therefore, as desired, $$\omega = p + (1 - p)\lr{1 + \frac{n - (qn + (q + 1)r)}{n(n - 1)}} = p + (1 - p)\lr{1 - \frac{(q - 1)n + (q + 1)r}{n(n - 1)}}.$$
\end{proof}  

\subsection{Classical Value}~\label{sub:classicalvalue}
We now prove that the synchronous strategy yielded by Theorem \ref{thm:ourclassicalstrategymu} is optimal among all classical strategies if the probability $p$ that Alice and Bob get equal inputs is at least $\frac{2}{m + 2}$.

Recall that we defined $s := \frac{p(n - 1)}{1 - p} + 1$, and that $s$ is monotonic in $p$ by Lemma $\ref{lem:smonotone}$.

\begin{mylemma}~\label{lem:classicalsynchoptimals}
    Classically, a synchronous strategy is optimal when $s = 2\left\lfloor\frac{n - 1}{m}\right\rfloor + 1$.
\end{mylemma}

\begin{proof}
    By Lemmas \ref{lem:mukatleastxikssquared} and \ref{lem:xikssquaredminimum}, $$\mu \ge \sum_{k = 0}^{m - 1}\xi_k^2 \ge \left\lfloor\frac{\xi}{m}\right\rfloor^2(m - (\xi \bmod m)) + \lr{\left\lfloor\frac{\xi}{m}\right\rfloor + 1}^2(\xi \bmod m).$$ 
    If $m \mid \xi$, we have $\left\lfloor\frac{\xi}{m}\right\rfloor = \left\lfloor\frac{\xi - 1}{m}\right\rfloor + 1$, so reducing $\xi$ by $1$ decreases the bound on $\mu$ by $$\left\lfloor\frac{\xi}{m}\right\rfloor^2 - \lr{\left\lfloor\frac{\xi}{m}\right\rfloor - 1}^2 = 2\left\lfloor\frac{\xi}{m}\right\rfloor - 1 = 2\left\lfloor\frac{\xi - 1}{m}\right\rfloor + 1.$$
    Otherwise, $\left\lfloor\frac{\xi}{m}\right\rfloor = \left\lfloor\frac{\xi - 1}{m}\right\rfloor$, so reducing $\xi$ by $1$ again decreases the bound on $\mu$ by $$\lr{\left\lfloor\frac{\xi}{m}\right\rfloor + 1}^2 - \left\lfloor\frac{\xi}{m}\right\rfloor^2 
    = 2\left\lfloor\frac{\xi}{m}\right\rfloor + 1
    = 2\left\lfloor\frac{\xi - 1}{m}\right\rfloor + 1.$$
    Furthermore, since $2\left\lfloor\frac{\xi - 1}{m}\right\rfloor + 1$ is (nonstrictly) monotonic in $\xi$, by induction on $\xi$, reducing $\xi$ by any amount $h \in \Z^+$ decreases the bound on $\mu$ by at most $h\lr{2\left\lfloor\frac{\xi - 1}{m}\right\rfloor + 1}$.

    By Theorem \ref{thm:ourclassicalstrategymu}, $\mu = q^2(m - r) + (q + 1)^2r$ is attained by some synchronous strategy. 

    Therefore, $\lr{2\left\lfloor\frac{n - 1}{m}\right\rfloor + 1}(n - \xi) \ge q^2(m - r) + (q + 1)^2r - \mu$, so $$\lr{2\left\lfloor\frac{n - 1}{m}\right\rfloor + 1}n - (q^2(m - r) + (q + 1)^2r) \ge \lr{2\left\lfloor\frac{n - 1}{m}\right\rfloor + 1}\xi - \mu:$$ the synchronous strategy has at least as large a value of $\nu := s\xi - \mu$, and hence at least as large a value $\omega$, as any other strategy at $s = 2\left\lfloor\frac{n - 1}{m}\right\rfloor + 1$.
\end{proof}

\begin{mytheorem}~\label{thm:classicalsynchoptimalp}
    Classically, a synchronous strategy is optimal for all $p \ge \frac{2}{m + 2}$.
\end{mytheorem}

\begin{proof}
    Since $n > 1$ and $\left\lfloor\frac{n - 1}{m}\right\rfloor \leq \frac{n - 1}{m}$, the value of $p$ corresponding to $s = 2\left\lfloor\frac{n - 1}{m}\right\rfloor + 1$ is $$p = \frac{\lr{2\left\lfloor\frac{n - 1}{m}\right\rfloor + 1} - 1}{n + \lr{2\left\lfloor\frac{n - 1}{m}\right\rfloor + 1} - 2} = \frac{2\left\lfloor\frac{n - 1}{m}\right\rfloor}{n + 2\left\lfloor\frac{n - 1}{m}\right\rfloor - 1} \le \frac{2\frac{n - 1}{m}}{n - 1 + 2\frac{n - 1}{m}} = \frac{2}{m + 2}.$$
    By Lemma \ref{lem:classicalsynchoptimals}, a synchronous strategy is optimal at $s = 2\left\lfloor\frac{n - 1}{m}\right\rfloor + 1$ and thus at some $p \le \frac{2}{m + 2}$. Therefore, by Theorem \ref{thm:synchoptimalatpmeansoptimalabovep}, it is optimal for all $p \ge \frac{2}{m + 2}$, as desired.
\end{proof}

\begin{mytheorem}~\label{thm:classicalvalue}
    When $p \ge \frac{2}{m + 2}$, the classical value of this game is $$p + (1 - p)\lr{1 - \frac{(q - 1)n + (q + 1)r}{n(n - 1)}}.$$
\end{mytheorem}
\begin{proof}
By Theorem \ref{thm:classicalsynchoptimalp}, the strategy defined in \S\ref{sub:ourclassicalstrategy} is optimal among classical strategies. 
    
    Then our game's value equals the value of this strategy, which by Theorem~\ref{thm:ourclassicalstrategyvalue} is $$p + (1 - p)\lr{1 - \frac{(q - 1)n + (q + 1)r}{n(n - 1)}}.$$
\end{proof}

\subsection{Other Classical Strategies}
Having shown that a synchronous strategy is optimal whenever $p \ge \frac{2}{m + 2}$, we now turn our attention to determining which classical strategies could be optimal when $p$ is small. 

Recall that $a_k := |\{x \mid a(x) = k\}|$, $b_k := |\{y \mid b(y) = k\}|$, and $\xi_k := |\{x \mid a(x) = b(x) = k\}|$. 
\begin{mylemma}\label{lem:xiatmostsumofminab}     For any fixed valid set of values $a_k$ and $b_k$, there exists a strategy that attains $\xi = \sum_{k = 0}^{m - 1} \min(a_k, b_k)$, and any such strategy must have $\xi_k = \min(a_k, b_k)$ for every $k \in [m]$.
\end{mylemma}

\begin{proof}
    Alice and Bob iterate together through all $k \in [m]$, keeping track of which inputs (in sorted order) they have allocated. They each allocate the first $\min(a_k, b_k)$ of their remaining inputs to output $k$, which by induction will always be the same $\min(a_k, b_k)$ inputs. They then allocate the last $a_k - \min(a_k, b_k)$ and $b_k - \min(a_k, b_k)$, respectively, of their remaining inputs to output $k$; at most one of these quantities will be nonzero. Hence, $\xi_k = \min(a_k, b_k)$.
    
    Alice and Bob then agree on the first $\sum_{k = 0}^{m - 1} \min(a_k, b_k)$ inputs, so $\xi = \sum_{k = 0}^{m - 1} \min(a_k, b_k)$.
    
    Furthermore, Alice allocates exactly $\min(a_k, b_k) + a_k - \min(a_k, b_k) = a_k$ inputs to each output $k$, and likewise Bob allocates exactly $\min(a_k, b_k) + b_k - \min(a_k, b_k) = b_k$ inputs to each output $k$, as desired.
    Since $\sum_{k = 0}^{m - 1} a_k = \sum_{k = 0}^{m - 1} b_k = n$ and the players' pointers each meet in the middle, our algorithm will only allocate each input exactly once, as desired. 

    By Lemma \ref{lem:mukleakandbk}, $\xi_k \le a_k$ and $\xi_k \le b_k$, and thus $\xi_k \le \min(a_k, b_k)$, for each $k$. Therefore, any strategy with $\xi_k \neq \min(a_k, b_k)$ for any $k \in [m]$ must have $\xi < \sum_{k = 0}^{m - 1} \min(a_k, b_k)$.
\end{proof}

Since $\mu = \sum_{k = 0}^{m - 1} a_kb_k$ is fully determined by the values of the $a_k$ and $b_k$, any optimal strategy must maximize $\xi$ given its values of $a_k$ and $b_k$, and thus by Lemma \ref{lem:xiatmostsumofminab} must satisfy $\xi_k = \min(a_k, b_k)$ for every $k \in [m]$. 
The players must dump their remaining $n - \xi$ inputs not accounted for by the $\xi_k$ into disjoint subsets of $[m]$: if Alice and Bob both dumped an input into some output $k$, either $\xi_k$ would increase or $\xi_k \neq \min(a_k, b_k)$. Then the increase in $\mu = \sum_{k = 0}^{m - 1}a_kb_k$ that Alice or Bob incurs by dumping one of their inputs into some output $k$ must be exactly $\xi_k$, so
they minimize $\mu$ by dumping their respective remaining $n - \xi$ inputs into the outputs (hereafter, $u$ and $v$) with the least and second least values of $\xi_k$. 

Assuming now that $m \ge 3$, we have $\xi_u \le \xi_v \le \xi_k$ for all $k \notin \{u, v\}$, so
$$\mu 
= \sum_{k = 0}^{m - 1} \xi_k^2 - \xi_u^2 - \xi_v^2 + \xi_u(\xi_u + n - \xi) + \xi_v(\xi_v + n - \xi) 
= \sum_{k = 0}^{m - 1} \xi_k^2 + (n - \xi)(\xi_u + \xi_v).$$
Then for any fixed value of $\xi_u + \xi_v$, we can minimize $\mu$ by separately minimizing $\xi_u^2 + \xi_v^2$ (subject to the fixed value of $\xi_u + \xi_v$) and $\sum_{k \notin\{u, v\}} \xi_k^2$ subject to $\sum_{k \notin\{u, v\}} \xi_k = \xi - (\xi_u + \xi_v)$.

By Lemma \ref{lem:xikssquaredminimizer} (with $m = 2$ and the fixed sum as $\xi_u + \xi_v$), $\xi_u^2 + \xi_v^2$ is minimized when $\xi_u = \left\lfloor\frac{\xi_u + \xi_v}{2}\right\rfloor$, and $\xi_v = \left\lfloor\frac{\xi_u + \xi_v}{2}\right\rfloor$ if $\xi_u + \xi_v$ is even and $\xi_v = \left\lfloor\frac{\xi_u + \xi_v}{2}\right\rfloor + 1$ otherwise.

Similarly, by Lemma \ref{lem:xikssquaredminimizer} on the remaining $m - 2$ terms (with the fixed sum as $\xi - (\xi_u + \xi_v)$), $\sum_{k \notin\{u, v\}} \xi_k^2$ is minimized when $\xi_k = \left\lfloor\frac{\xi - (\xi_u + \xi_v)}{m - 2}\right\rfloor + 1$ for exactly $(\xi - (\xi_u + \xi_v)) \bmod (m - 2)$ values of $k$ and $\xi_k = \left\lfloor\frac{\xi - (\xi_u + \xi_v)}{m - 2}\right\rfloor$ for the rest.

Given values for $\xi$ and $\xi_u + \xi_v$, we then know the optimal strategy. 

Since $\xi_u \le \xi_k$ for all $k \in [m]$, $m\xi_u = \sum_{k = 0}^{m - 1} \xi_u \le \sum_{k = 0}^{m - 1} \xi_k = \xi$  and thus $\xi_u \le \frac{\xi}{m}$. Also, in order to be optimal, we must have $\xi_v \le \left\lfloor\frac{\xi_u + \xi_v}{2}\right\rfloor + 1 \le \xi_u + 1$, so $\xi_u + \xi_v \le 2\xi_u + 1 \le 2\frac{\xi}{m} + 1$. 

Therefore, for any $(n, m)$, the number of strategies that might be optima at any $p$ is at most $$\sum_{\xi = 0}^n \lr{2\frac{\xi}{m} + 2} = \frac{2}{m}\lr{\sum_{\xi = 0}^n \xi} + 2(n + 1) = \frac{2}{m} \cdot \frac{n(n + 1)}{2} + 2(n + 1) = \lr{\frac{n}{m} + 2}(n + 1).$$ 
We can then iterate through these strategies in time $O\lr{\frac{n^2}{m}}$, 
and computing each of these strategies' values of $\nu$ requires $O(1)$ arithmetic operations, 
so finding the optimal classical strategy at a given value of $p$ is computationally efficient for $m \ge 3$. 

Computing the value of $p$ at which each such strategy's value intersects (i.e., equals) that of the synchronous strategy from \S\ref{sub:ourclassicalstrategy} also uses $O(1)$ arithmetic operations, so we can find the threshold for $p$ at which the synchronous strategy becomes optimal in $O\lr{\frac{n^2}{m}}$ time as well. 

\section{Quantum Strategies}

We now consider strategies where Alice and Bob are allowed to share an arbitrarily large finite-dimensional entangled quantum state before receiving inputs; they still may not communicate once they receive their inputs. We use the definition found in definition 4.1 of \cite{rnoqsfngtsuames}, where a quantum strategy consists of a bipartite pure state shared between Alice and Bob, who each apply a different complete set of projective measurements (PVMs) for each input. 

In this section, we will present a synchronous quantum strategy for our coloring game. We will refine this strategy to a more elegant form that highlights that the entanglement seems to be used to account for the fact that $m$ fails to divide $n$ (which precludes Alice and Bob from classically assigning the same number of inputs to each output). We will then show that strategy is optimal across quantum synchronous strategies, as well as across all quantum strategies when $p$ (the probability that the players receive equal inputs) is at least $\frac{2n-m}{(m+2)n-2m}$.

\subsection{An Optimal Synchronous Quantum Strategy}~\label{sub:ourquantumstrategy}
Recall the definition of a synchronous strategy: Alice and Bob must always give the same output when given the same input. Note that this restriction does not imply that they have a fixed output for each input, only that their outputs are equal if their inputs were equal. 
Let $D \in \C^{m \times m}$ be the diagonal matrix with $D_{kk} = \omega_n^k := e^{2\pi i k/n}$ for all $k \in [m]$.

Let $F$ be the $m \times m$ Fourier matrix: i.e., $F_{kl} = \frac{\omega_m^{kl}}{\sqrt{m}}$ where $\omega_m := e^{2\pi i/m}$. 

Alice and Bob initially have one half each of the $m$-dimensional Bell state $\frac{1}{\sqrt{m}}\sum_{k = 0}^{m - 1} \ket{k}\ket{k}$. 

Alice applies the unitary operator $F^*D^{-x}$ (where $U^*$ denotes the entrywise conjugate of $U$) upon receiving her input $x \in [n]$, and Bob applies $FD^{y}$ upon receiving input $y \in [n]$, after which both players measure in the standard basis and output their measurement result.

The state after they apply their respective operators is
$$
\begin{aligned}
& ((F^*D^{-x}) \otimes (FD^y))\lr{\frac{1}{\sqrt{m}}\sum_{k = 0}^{m - 1} \ket{k}\ket{k}} \\
& = \frac{1}{\sqrt{m}}\sum_{k = 0}^{m - 1} \lr{\lr{F^*D^{-x}\ket{k}} \otimes \lr{FD^y\ket{k}}} \\
& = \frac{1}{\sqrt{m}}\sum_{k = 0}^{m - 1} \lr{\lr{F^*(D_{kk})^{-x}\ket{k}} \otimes \lr{F(D_{kk})^{y}\ket{k}}} \\
& = \frac{1}{\sqrt{m}}\sum_{k = 0}^{m - 1} \lr{\lr{F^*(\omega_n^k)^{-x}\ket{k}} \otimes \lr{F(\omega_n^k)^{y}\ket{k}}} \\
& = \frac{1}{\sqrt{m}}\sum_{k = 0}^{m - 1} \lr{\lr{F^*\omega_n^{-kx}\ket{k}} \otimes \lr{F\omega_n^{ky}\ket{k}}} \\
& = \frac{1}{\sqrt{m}}\sum_{k = 0}^{m - 1} \omega_n^{ky - kx}\lr{\lr{F^*\ket{k}} \otimes \lr{F\ket{k}}}.
\end{aligned}
$$
The amplitude of any basis state $\ket{j}
\ket{j}$ with $j \in [m]$ is then
$$
\begin{aligned}
& \frac{1}{\sqrt{m}}\sum_{k = 0}^{m - 1} \omega_n^{ky - kx}\lr{(F^*\ket{k})_j}\lr{(F\ket{k})_j} \\
& = \frac{1}{\sqrt{m}}\sum_{k = 0}^{m - 1} \omega_n^{ky - kx}\lr{(F^*)_{jk}}\lr{(F)_{jk}} \\
& = \frac{1}{\sqrt{m}}\sum_{k = 0}^{m - 1} \omega_n^{ky - kx}\lr{\frac{\omega_m^{jk}}{\sqrt{m}}}^*\lr{\frac{\omega_m^{jk}}{\sqrt{m}}} \\
& = \frac{1}{\sqrt{m}}\sum_{k = 0}^{m - 1} \omega_n^{ky - kx}\lr{\frac{\omega_m^{-jk}}{\sqrt{m}}}\lr{\frac{\omega_m^{jk}}{\sqrt{m}}} \\
& = m^{-3/2}\sum_{k = 0}^{m - 1} \omega_n^{ky - kx} \omega_m^{-jk + jk} \\
& = m^{-3/2}\sum_{k = 0}^{m - 1} \omega_n^{ky - kx}\omega_m^{0} \\
& = m^{-3/2}\sum_{k = 0}^{m - 1} \omega_n^{k(y - x)}.
\end{aligned}
$$
The probability of measuring each such state is then 
$$
\begin{aligned}
& \lrb{m^{-3/2}\sum_{k = 0}^{m - 1} \omega_n^{k(y - x)}}^2 \\
& = \lr{m^{-3/2}\sum_{k = 0}^{m - 1} \omega_n^{k(y - x)}}^*\lr{m^{-3/2}\sum_{l = 0}^{m - 1} \omega_n^{l(y - x)}} \\
& = m^{-3}\lr{\sum_{k = 0}^{m - 1} \lr{\omega_n^{k(y - x)}}^*}\lr{\sum_{l = 0}^{m - 1} \omega_n^{l(y - x)}} \\
& = m^{-3}\lr{\sum_{k = 0}^{m - 1} \omega_n^{-k(y - x)}}\lr{\sum_{l = 0}^{m - 1} \omega_n^{l(y - x)}} \\
& = m^{-3}\sum_{k = 0}^{m - 1}\sum_{l = 0}^{m - 1} \omega_n^{-k(y - x)}\omega_n^{l(y - x)} \\
& = m^{-3}\sum_{k = 0}^{m - 1}\sum_{l = 0}^{m - 1} \omega_n^{(l - k)(y - x)}.
\end{aligned}
$$
There is one such state for each $j \in [m]$, so the probability of measuring any one of them is $$\sum_{j = 0}^{m - 1} m^{-3}\sum_{k = 0}^{m - 1}\sum_{l = 0}^{m - 1} \omega_n^{(l - k)(y - x)} = m \cdot m^{-3}\sum_{k = 0}^{m - 1}\sum_{l = 0}^{m - 1} \omega_n^{(l - k)(y - x)} = m^{-2}\sum_{k = 0}^{m - 1}\sum_{l = 0}^{m - 1} \omega_n^{(l - k)(y - x)}.$$
\begin{mytheorem}~\label{thm:quantumstrategyissynchronous}
    The strategy defined in this section is synchronous (and hence $\xi = n$).
\end{mytheorem}
\begin{proof}
    When Alice and Bob receive equal inputs, $y = x$, so $y - x = 0$ and this sum becomes
    $$
    m^{-2}\sum_{k = 0}^{m - 1}\sum_{l = 0}^{m - 1} \omega_n^{(l - k)0} 
    = m^{-2}\sum_{k = 0}^{m - 1}\sum_{l = 0}^{m - 1} \omega_n^{0} 
     =m^{-2}\sum_{k = 0}^{m - 1}\sum_{l = 0}^{m - 1} 1 
    = m^{-2}\sum_{k = 0}^{m - 1} m 
    = m^{-2} \cdot m^2
    = 1.
    $$
        Hence, Alice and Bob always win when $x = y$, so the strategy is synchronous, as desired.
\end{proof}

Recall that we defined $\mu$ for a given strategy as the sum across all input pairs $(x, y) \in [n]^2$ of the probabilities that the players give equal outputs.

\begin{mytheorem}~\label{thm:ourquantumstrategymu}
    The strategy defined in this section has $\mu = \frac{n^2}{m}$.
\end{mytheorem}
\begin{proof} We sum the probability of measuring equal values across all $x$ and $y$ to get
    $$\mu 
    = \sum_{x = 0}^{n - 1}\sum_{y = 0}^{n - 1} m^{-2}\sum_{k = 0}^{m - 1}\sum_{l = 0}^{m - 1} \omega_n^{(l - k)(y - x)} 
    = m^{-2}\sum_{x = 0}^{n - 1}\sum_{k = 0}^{m - 1}\sum_{l = 0}^{m - 1}\omega_n^{-x(l - k)}\sum_{y = 0}^{n - 1} \omega_n^{y(l - k)}.$$
    If $l = k$, the terms in the innermost sum fail to interfere:
    $$
    \begin{aligned}
    & \omega_n^{-x(l - k)}\sum_{y = 0}^{n - 1} \omega_n^{y(l - k)} = \omega_n^{-x \cdot 0}\sum_{y = 0}^{n - 1} \omega_n^{y \cdot 0} = \omega_n^{0}\sum_{y = 0}^{n - 1} \omega_n^{0} = 1\sum_{y = 0}^{n - 1} 1 = 1 \cdot n = n.
    \end{aligned}
    $$
    Otherwise, $l \not\equiv k \pmod{n}$ (since $0 \le l, k < m$ and thus $|l - k| < m \le n$), so $\omega_n^{l - k} \neq 1$ and the terms will interfere. By the standard formula for the sum of a finite geometric series,
    $$
    \sum_{y = 0}^{n - 1} \omega_n^{y(l - k)} 
    = \sum_{y = 0}^{n - 1} (\omega_n^{l - k})^{y} 
    = \frac{1 - (\omega_n^{l - k})^n}{1 - \omega_n^{l - k}} 
    = \frac{1 - (\omega_n^n)^{l - k}}{1 - \omega_n^{l - k}}
    = \frac{1 - 1^{l - k}}{1 - \omega_n^{l - k}}
    = \frac{1 - 1}{1 - \omega_n^{l - k}}
    = 0.
    $$
    Note that the denominator is $0$ (and hence the calculation fails) exactly when $\omega_n^{l - k} = 1$.

    Hence, for each $k$ exactly one $l$ (i.e., $l = k$) yields a sum of $n$ and the others yield $0$, so $$\sum_{l = 0}^{m - 1}\omega_n^{-x(l - k)}\sum_{y = 0}^{n - 1} \omega_n^{y(l - k)} = 1 \cdot n + (m - 1) \cdot 0 = n$$ for all $k \in [m]$ and all $x \in [n]$, and hence, as desired, $$\mu = m^{-2}\sum_{x = 0}^{n - 1}\sum_{k = 0}^{m - 1}\sum_{l = 0}^{m - 1}\omega_n^{-x(l - k)}\sum_{y = 0}^{n - 1} \omega_n^{y(l - k)} = m^{-2}\sum_{x = 0}^{n - 1}\sum_{k = 0}^{m - 1} n = m^{-2} \cdot n \cdot m \cdot n = \frac{n^2}{m}.$$
\end{proof}

Recall that we defined $p$ as the probability that the players receive equal inputs. 

\begin{mytheorem}~\label{thm:quantumstrategyvalue}
    The value of the strategy defined in this section is $p + (1 - p)\lr{1 - \frac{n - m}{m(n - 1)}}$.
\end{mytheorem}

\begin{proof}
    Recall that we can write the strategy's value in terms of $\xi$ and $\mu$: $$
    \omega = 1 - p + \lr{\frac{p}{n} + \frac{1 - p}{n(n - 1)}}\xi - \frac{1 - p}{n(n - 1)}\mu.$$
    By Theorem \ref{thm:quantumstrategyissynchronous}, $\xi = n$, and by Theorem \ref{thm:ourquantumstrategymu}, $\mu = \frac{n^2}{m}$, giving $$\begin{aligned}
    \omega & = 1 - p + \lr{\frac{p}{n} + \frac{1 - p}{n(n - 1)}}n - \frac{1 - p}{n(n - 1)}\frac{n^2}{m} \\
    & = 1 - p + p + \frac{1 - p}{n - 1} - \frac{(1 - p)n}{(n - 1)m} \\
    & = p + (1 - p)\lr{1 + \frac{1}{n - 1} - \frac{n}{(n - 1)m} } \\
    & = p + (1 - p)\lr{1 - \frac{n - m}{(n - 1)m}}.
    \end{aligned}$$
\end{proof}

\subsection{Using Fewer Measurement Bases}~\label{subsec:fewermeasurementbases}
When $n \ge 2m$, we can attain the same success probability using a strategy that drastically reduces the number of distinct measurement operations Alice and Bob must be able to apply.

For any $k \ge 0$ and $n \ge m$, consider the case of our game with $n + km$ inputs and $m$ outputs. 

\begin{mytheorem}~\label{thm:fewerbasesstrathassamevalue}
     The strategy where Alice and Bob each play the quantum strategy from \S\ref{sub:ourquantumstrategy} for the $(n, m)$ game on the first $n$ inputs and assign the remaining $km$ inputs evenly and identically across $[m]$  has the same value as the strategy from \S\ref{sub:ourquantumstrategy} for the $(n + km, m)$ game.
\end{mytheorem}

\begin{proof}

This strategy is synchronous on the first $n$ inputs (by Theorem \ref{thm:quantumstrategyissynchronous}) and the other $km$ inputs (since the players assign inputs identically), and hence is synchronous overall.

When both players' inputs are distinct and among the first $n$, their losing probability is $\frac{(n - m)}{m(n - 1)}$, given by that of the quantum strategy, specifically the complement of the coefficient of $1 - p$ in the quantum strategy's value (as computed in Theorem \ref{thm:quantumstrategyvalue}).

If only one input is among the first $n$, the inputs must be distinct; for each measurement outcome, there are exactly $k$ inputs (out of $km$) the other player could receive that would result in equal outputs (and hence a loss), so the losing probability in this case is $\frac{k}{km} = \frac{1}{m}$.

Finally, for each of Alice's $km$ inputs not among the first $n$, there are $k - 1$ inputs for Bob not among the first $n$ or equal to Alice's that result in equal outputs (and hence a loss).

Thus, when given different inputs, we lose with probability 
\begin{align*}
        & \frac{n(n - 1)\frac{n - m}{m(n - 1)} + 2nkm\frac{1}{m} + km(k - 1)}{(n + km)(n + km - 1)} \\
    = & \frac{n(n - m) + m(2nk + km(k - 1))}{m(n + km)(n + km - 1)} \\
    = & \frac{n^2 - nm + 2nkm + km^2(k - 1)}{m(n + km)(n + km - 1)} \\
    = & \frac{n^2 - nm + nkm + nkm + k^2m^2 - km^2}{m(n + km)(n + km - 1)} \\
    = & \frac{n^2 + nkm - nm + nkm + k^2m^2 - km^2}{m(n + km)(n + km - 1)} \\
    = & \frac{n(n + km - m) + km(n + km - m)}{m(n + km)(n + km - 1)} \\
    = & \frac{(n + km)(n + km - m)}{m(n + km)(n + km - 1)} \\
    = & \frac{(n + km - m)}{m(n + km - 1)},
\end{align*}
exactly the losing probability of the optimal quantum strategy for $(n + km, m)$, as desired.
\end{proof}

An immediate corollary is that it is never necessary to actually play the quantum strategy for any $n \ge 2m$: we could instead just play the smaller quantum strategy on the first $m + (n \bmod m)$ inputs and the classical strategy on the rest. We refer to this as the ``mod $m$'' strategy.

\subsection{Quantum Synchronous Value}~\label{sub:quantumsynchronousoptimality}
We now prove that the quantum value our strategy achieves is optimal across all synchronous strategies (i.e., ones where Alice and Bob give the same output when given the same input). 

\begin{myremark}
    Synchronous strategies must always win this game on equal inputs. Thus, to prove that a synchronous strategy is optimal across all synchronous strategies for this game on the input distributions we consider, it suffices to prove that said strategy is optimal when inputs are always unequal: i.e., $p=0$.
\end{myremark}

\begin{mytheorem}~\label{thm:quantumsynchronousoptimality}
When the players always receive unequal inputs, no synchronous quantum strategy has a value greater than that of the strategy from \S\ref{sub:ourquantumstrategy}, which is $$1-\frac{n-m}{m(n-1)}.$$
\end{mytheorem}
\begin{proof}
    This relies on two theorems found in 
        ~\cite{svog}. The first, Theorem 3.6, gives us that the synchronous quantum value of an $m$-coloring game on a graph $G(V,E)$ with the input distribution being uniform over all unequal inputs is at most $$1-\frac{m}{|E|}f_{G,q}\lr{\frac{1}{m}}$$where $f_{G,q}$ is the quantum graph correlation function ~\cite{svog}.
        Theorem 3.8 of ~\cite{svog} gives us that $f_{K_n,q}(t) = nt(nt-1)$ for all $n \geq 5$, as long as $$\frac{n-\sqrt{n^2-4n}}{2n} \le t \leq \frac{n+\sqrt{n^2-4n}}{2n}.$$
Plugging in our value of $\frac{1}{m}$, we find that, when the above constraints are met, the quantum synchronous value of the graph coloring game with $m$ colors played on $K_n$ with the input distribution being uniform over all unequal inputs is at most $$1-\frac{m}{n^2-n}\lr{\frac{n}{m}\lr{\frac{n}{m}-1}}$$which simplifies to $$1-\frac{1}{n-1}\lr{\frac{n}{m}-1} = 1-\frac{n-m}{m(n-1)}$$exactly the value of our quantum strategy. Thus, our synchronous quantum strategy is optimal whenever $$n \geq 5, \frac{n-\sqrt{n^2-4n}}{2n} \le \frac{1}{m} \leq \frac{n+\sqrt{n^2-4n}}{2n}$$But we can do better.

Considering only interesting cases ($n > m \ge 2$), the $n \geq 5$ requirement leaves us with three parameter pairs where our strategy may not be optimal, which we write in the form $(n,m)$: $(3,2), (4,2), (4,3)$. The second condition is more problematic. Since $m \geq 2$, we know that $$\frac{1}{m} \leq \frac{1}{2} < \frac{n+\sqrt{n^2-4n}}{2n}$$A simple algebraic proof gives that, for all $m \leq n-2$, we have $$\frac{n-\sqrt{n^2-4n}}{2n} < \frac{1}{m}:$$ 

\begin{align*}& \frac{n-\sqrt{n^2-4n}}{2n} \cdot \frac{n+\sqrt{n^2-4n}}{n+\sqrt{n^2-4n}} \\
& = \frac{n^2-n^2+4n}{2n(n+\sqrt{n^2-4n})} \\
& = \frac{1}{\frac{1}{2}(n+\sqrt{n^2-4n})} \end{align*}

And so it suffices to prove that $$\frac{1}{2}(n+\sqrt{n^2-4n}) > n-2$$or, more simply that $$\sqrt{n^2-4n} > 2n-4-n = n-4$$From the other restriction we have that $n > 4$, so $4n > 16$ and thus $$-4n > -8n + 16$$ so $$n^2-4n > n^2 -8n + 16 = (n-4)^2$$We can simply take the square root of both sides to find that $$\sqrt{n^2-4n} > n-4$$

Thus, in addition to being unsure of optimality for parameters $(3,2),(4,2), $ and $(4,3)$, we also are unsure of the optimality of our strategy for all pairs of parameters $(n,n-1)$. Fortunately, the result from \S\ref{subsec:fewermeasurementbases} (specifically the mod $m$ strategy) may be used to prove optimality in these cases as well. Suppose for the sake of contradiction that there existed some synchronous quantum strategy with value greater than 
$$ 1-\frac{z-(z-1)}{(z-1)(z-1)}$$
for some parameter pair $(n=z,m=z-1)$. Note that, since the only interesting cases are when $m \geq 2$, we can assume $z \geq 3$. By the mod $m$ strategy, we know that, when playing the game with parameter pair $(n' = 2z-1, m' = z-1)$, the optimal success probability can be achieved by simply playing our quantum strategy on the first $z$ inputs and the classical strategy on the other $z-1$ inputs. If we instead played this superior quantum strategy on the first $z$ inputs and the classical strategy on the other $z-1$ inputs, then we would achieve a strictly higher success probability. This follows from the proof of Theorem \ref{thm:fewerbasesstrathassamevalue}: We have no guarantees on what outputs the hypothetical quantum strategy for $(z,z-1)$ will give, but since the number of inputs assigned to each output in the classical portion is $1$, we will still have the exact same average failure rate as the scenario in Theorem \ref{thm:fewerbasesstrathassamevalue} when one input is in the classically handled inputs ($z$ to $2z-2$) and one is in the quantum-handled inputs (0 to $z-1$). The failure rate in the both-inputs-handled-classically case should also be unchanged from the scenario in Theorem \ref{thm:fewerbasesstrathassamevalue}, so the only change is that, when both inputs are in the quantum-handled range, we have a better success probability. Thus, if the $(n=z,m=z-1)$ case had a synchronous quantum strategy with value greater than 
$$1-\frac{n-m}{m(n-1)} = 1-\frac{z-(z-1)}{(z-1)(z-1)}$$then the $(n'=2z-1,m'=z-1)$ case would have a value greater than 
$$1-\frac{n'-m'}{m'(n'-1)} = 1-\frac{2z-1-(z-1)}{(z-1)(2z-2)} $$
which we know reaches the upper bound, and thus we have a contradiction. In particular, this covers $(3,2)$ and $(4,3)$, since better strategies for $(3,2)$ or $(4,3)$ would imply better strategies for $(5,2)$ and $(7,3)$ respectively, which both violate our upper bound and thus create contradictions. For $(4,2)$, we can again invoke Theorem \ref{thm:fewerbasesstrathassamevalue} to observe that, for $(6,2)$, playing the $(4,2)$ quantum strategy on the first 4 inputs and playing the classical strategy on the remaining 2 inputs is optimal, so by the same argument as above a better synchronous quantum strategy for $(4,2)$ implies a better synchronous quantum strategy for $(6,2)$, creating a contradiction.

Thus, our synchronous quantum value of $$1-\frac{n-m}{m(n-1)}$$is optimal for all $n > m \ge 2$.
\end{proof} 
\begin{myremark}
    Note that Theorem $\ref{thm:quantumsynchronousoptimality}$ implies the strategy laid out in $\S$\ref{sub:ourquantumstrategy} is optimal across all quantum synchronous strategies.
\end{myremark}

\begin{mytheorem}~\label{thm:quantumsynchronousvalue}
The quantum synchronous value of this game is $p+(1-p)\lr{1-\frac{(n-m)}{m(n-1)}}.$
\end{mytheorem}
\begin{proof}
Since synchronous strategies can only lose this game on unequal inputs, this follows directly from Theorem $\ref{thm:quantumsynchronousoptimality}$ providing the upper bound and Theorem $\ref{thm:quantumstrategyvalue}$ meeting that bound.
\end{proof}

\subsection{Quantum Value}~\label{sub:quantumvalue} 
We now prove that the synchronous strategy detailed in \S\ref{sub:ourquantumstrategy}
is optimal when the probability $p$ that the players receive equal inputs is sufficiently large: i.e., $p\geq \frac{2n-m}{(m+2)n-2m}$. 

Recall the following definitions: $c_{xy}$ is the probability that Alice's output ($a$) is equal to Bob's output ($b$) given that Alice received input $x$ and Bob received input $y$. We also defined

$$\xi := \sum_{x = 0}^{n - 1}c_{xx}$$ and $$\mu := \sum_{x = 0}^{n - 1} \sum_{y = 0}^{n - 1}c_{xy}.$$

Finally, recall that our input distribution $\pi$ was defined as follows: letting $p := \mathbb{P}[x = y]$, $\pi(x, x) = \frac{p}{n}$ for each $x \in [n]$ and $\pi(x, y) = \frac{1 - p}{n(n - 1)}$ whenever $x \neq y$. Note that we only care to study $p \in (0,1)$, since $p=0$ and $p=1$ both have trivial strategies that always win.

As shown in \cite{rnoqsfngtsuames}, through Naimark's dilation theorem and purification, considering only projective measurement strategies on shared pure states is sufficient to characterize the quantum value of a nonlocal game, assuming Alice and Bob are restricted to a finite-dimensional Hilbert space. 
As such, we need only consider strategies where, for given inputs $x$ and $y$ respectively, $\{A_{x}^{a}\}_{a},\{B_{y}^{b}\}_{b}$ are Alice's and Bob's projective measurements, such that Alice outputs color $a$ when observing $a$, and Bob outputs color $b$ when observing $b$. 
Let 
$$\ket\psi \in \mathcal{H} = \mathcal{H}_A \otimes \mathcal{H}_B$$
be the pure state shared by Alice and Bob, where $\mathcal{H}_A$ is the Hilbert space for the portion of the state Alice owns, and $\mathcal{H}_B$ is the Hilbert space for the portion of the state Bob owns.

We can see that the probability that outputs are equal for a given pair of inputs $(x,y)$ is 
$$c_{xy} = \sum_{k = 0}^{m - 1} \Pr(k,k \mid x,y) = \sum_{k = 0}^{m - 1} \bra{\psi} A_x^k \otimes B_y^k \ket{\psi}$$
and thus that 
$$\xi = \sum_{x=0}^{n-1} \sum_{k = 0}^{m - 1} \bra{\psi} A_x^k \otimes B_x^k \ket{\psi}$$
and
$$\mu = \sum_{x = 0}^{n - 1} \sum_{y = 0}^{n - 1} \sum_{k = 0}^{m - 1} \bra{\psi} A_x^k \otimes B_y^k \ket{\psi}.$$
We can also define three new operators:
$$N_A^k := \sum_{x=0}^{n-1} A^k_x,$$
$$N_B^k := \sum_{y=0}^{n-1} B^k_y,$$
$$\Xi_k := \sum_{x=0}^{n-1}A^k_x \otimes B^k_x.$$

\begin{mylemma}~\label{lem:operatorpsd}
For each color $k$, the block operator
$$\Gamma_k = \begin{bmatrix}
N_A^k \otimes I & \Xi_k \\
\Xi_k & I \otimes N_B^k
\end{bmatrix}
$$
is positive semidefinite.
\end{mylemma}
\begin{proof}
We can prove this by constructing $\Gamma_k$ as a sum of positive semidefinite matrices. Let 
$$T_x = \begin{bmatrix}
    A_x^k \otimes I & I \otimes B_x^k 
\end{bmatrix}.$$
Note that an operator times its conjugate transpose is positive semidefinite, so $T_x^{\dagger}T_x$ is positive semidefinite. We can write it out: 
$$T_x^{\dagger}T_x = \begin{bmatrix}
    (A_x^k \otimes I)^\dagger \\ 
    (I \otimes B_x^k)^\dagger 
\end{bmatrix}\begin{bmatrix}
    A_x^k \otimes I & I \otimes B_x^k
\end{bmatrix} 
= 
\begin{bmatrix}
     (A_x^k \otimes I)^\dagger  (A_x^k \otimes I) &  (A_x^k \otimes I)^\dagger(I \otimes B_x^k ) \\
     (I \otimes B_x^k)^\dagger(A_x^k \otimes I)&(I \otimes B_x^k)^\dagger (I \otimes B_x^k)
\end{bmatrix}$$
Note that $(A_x^k \otimes I)$ and $(I \otimes B_x^k)$ are projectors, so they are Hermitian and equal to their own square, so we can rewrite this as 
$$\begin{bmatrix}
     (A_x^k \otimes I)^\dagger  (A_x^k \otimes I) &  (A_x^k \otimes I)^\dagger(I \otimes B_x^k ) \\
     (I \otimes B_x^k)^\dagger(A_x^k \otimes I)&(I \otimes B_x^k)^\dagger (I \otimes B_x^k)
\end{bmatrix}
=\begin{bmatrix}
     (A_x^k)^2 \otimes I &  A_x^k \otimes B_x^k \\
     A_x^k \otimes B_x^k&I \otimes (B_x^k)^2
\end{bmatrix}
=\begin{bmatrix}
     A_x^k \otimes I &  A_x^k \otimes B_x^k \\
     A_x^k \otimes B_x^k&I \otimes B_x^k
\end{bmatrix}.
$$
Taking the sum of this expression for $T_x^{\dagger}T_x$ over all inputs $x$ gives
$$\sum_{x=0}^{n-1}\begin{bmatrix}
     A_x^k \otimes I & A_x^k \otimes B_x^k \\
     A_x^k \otimes B_x^k & I \otimes B_x^k
\end{bmatrix}
=
\begin{bmatrix}
     \sum_{x=0}^{n-1}A_x^k \otimes I & \sum_{x=0}^{n-1} A_x^k \otimes B_x^k \\
     \sum_{x=0}^{n-1}A_x^k \otimes B_x^k & \sum_{x=0}^{n-1}I \otimes B_x^k
\end{bmatrix}
=
\begin{bmatrix}
N_A^k \otimes I & \Xi_k \\
\Xi_k & I \otimes N_B^k
\end{bmatrix}
= \Gamma_k$$ 
Thus, since $\Gamma_k$ is a sum of $n$ positive semidefinite matrices, it is positive semidefinite.
\end{proof}
Recall that the Löwner operator relation $Q \preceq R$ means that $R-Q$ is positive semidefinite.
\begin{mylemma}~\label{lem:operatorinequality}
For each color $k$, $\Xi_k \preceq (N_A^k)^{\frac{1}{2}} \otimes (N_B^k)^{\frac{1}{2}}$.
\end{mylemma}
\begin{proof}
Note: To reduce confusion, we will make explicit which Hilbert space each identity matrix acts on. We fix $\epsilon > 0$ and define the following operators:
$$Q := (N_A^k \otimes I_{\mathcal{H}_B}) + \epsilon I_{\mathcal{H}_A \otimes \mathcal{H}_B}  = (N_A^k + \epsilon I_{\mathcal{H}_A}) \otimes I_{\mathcal{H}_B}$$
$$R := (I_{\mathcal{H}_A} \otimes N_B^k) +\epsilon I_{\mathcal{H}_A \otimes \mathcal{H}_B}  = I_{\mathcal{H}_A} \otimes (N_B^k + \epsilon I_{\mathcal{H}_B}) $$
Note that $N_A^k$ and $N_B^k$ are sums of projectors and thus are positive semidefinite and Hermitian, so $N_A^k \otimes I_{\mathcal{H}_B}$ and $I_{\mathcal{H}_A} \otimes N_B^k$ are Hermitian and positive semidefinite and thus have non-negative eigenvalues. Thus, $Q$ and $R$ are Hermitian and will have strictly positive eigenvalues, and are thus positive definite. Also note that $Q$ and $R$ commute.

We now invoke a result from \cite{ANDO1979203}: for positive definite matrices $L,M$, the greatest Hermitian matrix $Z$ such that 
$$\begin{bmatrix}
    L & Z \\
    Z & M
\end{bmatrix}$$
is positive semidefinite is the geometric mean $L \#M$ which, when $L$ and $M$ commute, can be written as  $(LM)^\frac{1}{2}$ or as $L^{\frac{1}{2}}M^{\frac{1}{2}}$ (see \cite[p.~106]{bhatia}). By Lemma \ref{lem:operatorpsd}, 
$$\begin{bmatrix}
N_A^k \otimes I_{\mathcal{H}_B} & \Xi_k \\
\Xi_k & I_{\mathcal{H}_A} \otimes N_B^k
\end{bmatrix}$$
is positive semidefinite, so
$$\begin{bmatrix}
N_A^k \otimes I_{\mathcal{H}_B} & \Xi_k \\
\Xi_k & I_{\mathcal{H}_A} \otimes N_B^k
\end{bmatrix} + \epsilon\begin{bmatrix}
I_{\mathcal{H}_{A} \otimes \mathcal{H}_B} & 0 \\
0& I_{\mathcal{H}_{A} \otimes \mathcal{H}_B}
\end{bmatrix} $$
is also positive semidefinite. That matrix can be rewritten as 
$$\begin{bmatrix}
(N_A^k + \epsilon I_{\mathcal{H}_A}) \otimes I_{\mathcal{H}_B} & \Xi_k \\
\Xi_k & I_{\mathcal{H}_A} \otimes (N_B^k + \epsilon I_{\mathcal{H}_B})
\end{bmatrix} =\begin{bmatrix}
Q  & \Xi_k \\
\Xi_k & R
\end{bmatrix}$$
Thus, since $Q$ and $R$ commute and $\Xi_k$ is Hermitian (as a sum of tensor products of projectors),
$$\Xi_k \preceq Q^{
\frac{1}{2}}R^\frac{1}{2} 
= ((N_A^k + \epsilon I_{\mathcal{H}_A}) \otimes I_{\mathcal{H}_B})^{\frac{1}{2}}(I_{\mathcal{H}_A} \otimes (N_B^k + \epsilon I_{\mathcal{H}_B}))^{\frac{1}{2}} 
= (N_A^k + \epsilon I_{\mathcal{H}_A})^{\frac{1}{2}} \otimes (N_B^k + \epsilon I_{\mathcal{H}_B})^{\frac{1}{2}}.$$
Note that $N_A^k + \epsilon I_{\mathcal{H}_A}$ can be diagonalized using the same unitary as $N_A^k$, 
so we can write
$$N_A^k = U\mathrm{diag}(\lambda_1, ..., \lambda_d)U^\dagger$$
as well as
$$(N_A^k)^\frac{1}{2} = U\mathrm{diag}(\sqrt{\lambda_1}, ..., \sqrt{\lambda_d})U^\dagger$$
and
$$N_A^k + \epsilon I_{\mathcal{H}_A} = U\mathrm{diag}(\lambda_1 +\epsilon, ..., \lambda_d +\epsilon)U^\dagger$$
and thus
$$(N_A^k + \epsilon I_{\mathcal{H}_A})^{\frac{1}{2}} = U\mathrm{diag}(\sqrt{\lambda_1 +\epsilon}, ..., \sqrt{\lambda_d +\epsilon})U^\dagger,$$
so since the square root function on real non-negative scalars is continuous (and of course, the eigenvalues of a positive semidefinite matrix are real and non-negative), we have that
$$\lim_{\epsilon \rightarrow 0}(N_A^k + \epsilon I_{\mathcal{H}_A})^{\frac{1}{2}} = (N_A^k)^\frac{1}{2}$$
and by the same logic
$$\lim_{\epsilon \rightarrow 0}(N_B^k + \epsilon I_{\mathcal{H}_B})^{\frac{1}{2}} = (N_B^k)^\frac{1}{2}.$$
Thus, since limits preserve the Löwner order
and distribute over tensor products (when each tensor factor converges individually),
we can send $\epsilon$ to 0 to conclude that
$$\Xi_k \preceq (N_A^k + 0I_{\mathcal{H}_A})^{\frac{1}{2}} \otimes (N_B^k + 0I_{\mathcal{H}_B})^{\frac{1}{2}} = (N_A^k)
^{\frac{1}{2}} \otimes (N_B^k)^{\frac{1}{2}}.$$
\end{proof}

\begin{mytheorem}~\label{thm:xisquaredinequality}
Any valid quantum strategy for this game has $\mu \geq \frac{\xi^2}{m}$.
\end{mytheorem}

\begin{proof}
We can define a couple of new terms:
$$\xi_k 
:= \bra{\psi}\Xi_k\ket{\psi} 
= \bra{\psi}\lr{\sum_{x=0}^{n-1}(A_x^k \otimes B_x^k)} \ket{\psi}
= \sum_{x=0}^{n-1} \bra{\psi} (A_x^k \otimes B_x^k) \ket{\psi} 
.$$
$$\mu_k 
:= \bra{\psi}  (N_A^k \otimes N_B^k) \ket{\psi} 
= \bra{\psi} \lr{\lr{\sum_{x = 0}^{n - 1}A_x^k} \otimes \lr{\sum_{y = 0}^{n - 1}B_y^k}} \ket{\psi}
= \sum_{x = 0}^{n - 1} \sum_{y = 0}^{n - 1} \bra{\psi} (A_x^k \otimes B_y^k) \ket{\psi} 
.$$
Note from the formulas for $\xi$ and $\mu$ that 
$\xi = \sum_{k = 0}^{m - 1}\xi_k$
and
$\mu = \sum_{k = 0}^{m - 1}\mu_k$. 
We can then define 
$$R_k := (N_A^k)
^{\frac{1}{2}} \otimes (N_B^k)^{\frac{1}{2}}.$$
Note that $R_k$ is a tensor product of Hermitian operators (positive semidefinite Hermitian operators have positive semidefinite square roots) and thus is itself Hermitian, and
$$R_k^2 = (N_A^k)
^{\frac{1}{2}}(N_A^k)
^{\frac{1}{2}} \otimes (N_B^k)^{\frac{1}{2}}(N_B^k)^{\frac{1}{2}} = N_A^k \otimes N_B^k$$
and thus 
$$\mu_k =\bra{\psi}  R_k^2 \ket{\psi}.$$
By Lemma \ref{lem:operatorinequality}, 
$R_k \succeq \Xi_k$,
and thus we have that
$$\xi_k = \bra{\psi}\Xi_k\ket{\psi} \leq \bra{\psi}R_k\ket{\psi}.$$
We can observe that, since $R_k$ is Hermitian and thus $\braket{\psi \mid R_k\psi}$ is real, 
$$\braket{\psi \mid R_k\psi} \leq |\braket{\psi \mid R_k\psi}|$$
and by the Cauchy-Schwarz inequality on vectors in $\mathcal{H}_A \otimes \mathcal{H}_B$,
$$|\braket{\psi \mid R_k\psi}|\leq 
||\ket{\psi}|| \cdot ||\ket{R_k\psi}||.$$
The magnitude of $\ket{\psi}$ is 1, so 
$$||\ket{\psi}|| \cdot ||\ket{R_k\psi}||= ||\ket{R_k\psi}|| = || R_k \ket{\psi}||.$$
Thus
$$\xi_k  \leq || R_k \ket{\psi}||.$$
Since $R_k$ is Hermitian, 
$$|| R_k \ket{\psi}||^2 = \bra{\psi} R_k^\dagger R_k\ket{\psi} = \bra{\psi}R_k^2\ket{\psi} = \mu_k$$
and thus (since $\xi_k$ is a sum of probabilities and thus is nonnegative)
$$\xi_k^2 \leq \mu_k.$$
Then by the Cauchy-Schwarz inequality on $\mathbb{R}^m$,
$$\xi^2 = \lr{\sum_{k = 0}^{m - 1} \xi_k}^2 \leq \lr{\sum_{k = 0}^{m - 1} 1^2}\lr{\sum_{k = 0}^{m - 1} \xi_k^2} = m\sum_{k = 0}^{m - 1} \xi_k^2 
\leq m\sum_{k = 0}^{m - 1} \mu_k = m\mu $$
and thus $\frac{\xi^2}{m} \leq\mu$ as desired.
\end{proof}

Recall that we defined $s := \frac{p(n - 1)}{1 - p} + 1$, and that $s$ is monotonic in $p$ by Lemma \ref{lem:smonotone}.

\begin{mylemma}~\label{lem:quantumoptimals}
At $s = \frac{2n}{m}$, the strategy defined in  \S\ref{sub:ourquantumstrategy} is optimal among quantum strategies.
\end{mylemma}
\begin{proof}
    Recall that we can use $\nu := s\xi - \mu$ in lieu of the value $\omega$ as an objective function.     
    By Theorem \ref{thm:xisquaredinequality}, any valid quantum strategy has $\mu \ge \frac{\xi^2}{m}$, and thus $$\nu := s\xi - \mu = \frac{2n}{m}\xi - \mu  \le \frac{2n}{m}\xi - \frac{\xi^2}{m} = \frac{\xi(2n - \xi)}{m}.$$
    This expression is quadratic in $\xi$ with negative leading coefficient and hence is maximized when $\xi$ equals the midpoint of its roots: i.e., at $\xi = \frac{0 + 2n}{2} = n$. 
    
    The strategy defined in \S\ref{sub:ourquantumstrategy} attains both $\xi = n$ by Theorem \ref{thm:quantumstrategyissynchronous} and $\mu = \frac{n^2}{m} = \frac{\xi^2}{m}$ by Theorem \ref{thm:ourquantumstrategymu},     so no other quantum strategy can have a larger value of $\nu$ than the one achieved by the strategy defined in \S\ref{sub:ourquantumstrategy}, as desired.
\end{proof}

\begin{mytheorem}~\label{thm:quantumoptimalp}
For all $p \geq \frac{2n-m}{(m+2)n-2m}$, the quantum strategy defined in \S\ref{sub:ourquantumstrategy} is optimal among quantum strategies.
\end{mytheorem}
\begin{proof}
    By Lemma \ref{lem:quantumoptimals}, this strategy is optimal at the value of $p$ corresponding to $s = \frac{2n}{m}$: $$p = \frac{s - 1}{n + s - 2} = \frac{\frac{2n}{m} - 1}{n + \frac{2n}{m} - 2} = \frac{2n - m}{nm + 2n - 2m}.$$
    By Theorem \ref{thm:synchoptimalatpmeansoptimalabovep}, it is then optimal for all $p \geq \frac{2n-m}{(m+2)n-2m}$, as desired.
\end{proof}

\begin{mytheorem}~\label{thm:quantumgamevalue}
For all $p \geq \frac{2n-m}{(m+2)n-2m}$, our game has quantum value $$\omega = p + (1 - p)\lr{1 - \frac{n - m}{m(n - 1)}}.$$
\end{mytheorem}
\begin{proof}
    By Theorem \ref{thm:quantumoptimalp}, the strategy defined in \S\ref{sub:ourquantumstrategy} is optimal among quantum strategies. 
    
    Then its value, which by Theorem~\ref{thm:quantumstrategyvalue} is $p + (1 - p)\lr{1 - \frac{n - m}{m(n - 1)}}$, is the game's value.
\end{proof}

\begin{myremark}
    This parallels the classical threshold for synchronous optimality: when $m \ge 2$, $$\frac{2n-m}{(m+2)n-2m} \leq \frac{2}{m+2} \le \frac{1}{2}.$$
\end{myremark}
\begin{myremark}
Earlier in our work, we found a separate proof that $n\xi - \mu \leq n^2 (1-\frac{1}{m})$. This leads to a proof of the same value of this game which only holds when $p \geq \frac{1}{2}$. This is entirely covered by the above work, but since it was a substantially different proof technique, we have preserved that chain of work in Appendix \ref{app:quantumponehalf}.
\end{myremark}

\subsection{Other Quantum Strategies}
Having shown that a synchronous strategy is optimal whenever $p \ge \frac{2n-m}{(m+2)n-2m}$, we now try to find strategies that work better for smaller $p$. 
One obvious idea is for Alice to always output $0$ and Bob $1$, so that they win exactly when they receive different inputs: i.e., $\xi = \mu = 0$. 

Making use of the remaining $m - 2$ colors lets them do a bit better. If Alice's input $x$ satisfies $x < m - 2$, she can output $x$, and otherwise always output $m - 2$; likewise, if Bob's input $y$ satisfies $y < m - 2$, Bob can output color $y$, and otherwise always output $m - 1$. Then $\xi = m - 2$, and $\mu = \xi$ still, since Alice and Bob only output the same color if $x = y < m - 2$. 

In fact, if $m \ge 3$ they can play our optimal synchronous strategy (with $m - 2$ colors) on anywhere between $m - 2$ and $n$ of their inputs; as before, Alice can dump her remaining inputs into the second-to-last color and Bob can dump his into the last color.

Then $\xi$ equals the number of synchronous inputs exactly, and $\mu = \frac{\xi^2}{m - 2}$, so $$\nu = s\xi - \mu = s\xi - \frac{\xi^2}{m - 2}.$$ This is quadratic in $\xi$ with negative leading coefficient, so the value of $\xi$ that maximizes $\nu$ is simply the nearest element of $[m - 2, n] \cap \Z$ to $\frac{s(m - 2)}{2}$, which maximizes $\nu$ over $\R$.

In particular, if $\frac{s(m - 2)}{2} < m - 2$ (i.e., $s < 2$), the optimal $\xi$ is $m - 2$. Even if some analogue of these dumping strategies that allowed fractional $\xi$ were feasible (which we have not ruled out), $\xi < m - 2$ would imply that $\mu = \frac{\xi^2}{m - 2} = \xi\frac{\xi}{m - 2} < \xi$, which is impossible.

Likewise, if $\frac{s(m - 2)}{2} > n$ (i.e., $s > \frac{2n}{m - 2}$), the optimal $\xi$ is $n$ (even fractionally), and since this strategy is fully synchronous, it cannot outperform the strategy from \S\ref{sub:ourquantumstrategy} by Theorem \ref{thm:quantumsynchronousoptimality}.

The largest value of $\nu$ attained by any dumping strategy is then at most $$s\frac{s(m - 2)}{2} - \frac{\lr{\frac{s(m - 2)}{2}}^2}{m - 2} = \frac{s^2(m - 2)}{2} - \frac{s^2(m - 2)^2}{4(m - 2)} = \frac{s^2(m - 2)}{2} - \frac{s^2(m - 2)}{4} = \frac{s^2(m - 2)}{4}:$$
it could beat our synchronous strategy when $\frac{s^2(m - 2)}{4} > sn - \frac{n^2}{m}$, or $s^2m(m - 2) > 4smn - 4n^2$, which occurs exactly when the quadratic $m(m - 2)s^2 - 4mns + 4n^2$ is positive. Its roots are $$\frac{4mn \pm \sqrt{(4mn)^2 - 4 \cdot m(m - 2) \cdot 4n^2}}{2m(m - 2)} = \frac{4mn \pm 4n\sqrt{m^2 - m(m - 2)}}{2m(m - 2)} = \frac{2n(m \pm \sqrt{2m})}{m(m - 2)},$$ 
and its leading coefficient is positive, so it is positive when $s < \frac{2n(m - \sqrt{2m})}{m(m - 2)}$ or $s > \frac{2n(m + \sqrt{2m})}{m(m - 2)}$. 

We can ignore the latter case, since $\sqrt{2m} > 0$, so $s > \frac{2n(m + \sqrt{2m})}{m(m - 2)} > \frac{2nm}{m(m - 2)} = \frac{2n}{m - 2}$. 

The largest $s$ at which a dumping strategy could beat our synchronous strategy is then $$\frac{2n(m - \sqrt{2m})}{m(m - 2)},$$ as long as this quantity is greater than $2$, which occurs when $2n(m - \sqrt{2m}) > 2m(m - 2)$, or $2nm - 2m(m - 2) > 2n\sqrt{2m}$, which (since $2\sqrt{m} > 0$) is equivalent to $(n + 2 - m)\sqrt{m} > \sqrt{2}n$. Both sides are always positive, so this inequality is equivalent to $(n + 2 - m)^2m > 2n^2$, or $$
\begin{aligned}
    0 & < (n + 2 - m)^2m - 2n^2 \\
    & = (n^2 + 4 + m^2 + 4n - 2mn - 4m)m - 2n^2 \\
    & = mn^2 + 4m + m^3 + 4mn - 2m^2n - 4m^2 - 2n^2 \\
    & = mn^2  - 2n^2 + m^3 - 2m^2 - 2m^2n + 4mn - 2m^2 + 4m \\
    & = (m - 2)(n^2 + m^2 - 2mn - 2m) \\
    & = (m - 2)((n - m)^2 - 2m).
\end{aligned}$$ 
Since $m \ge 3$, $m - 2 > 0$, so this inequality simplifies to $0 < (n - m)^2 - 2m$, or $2m < (n - m)^2$.

Otherwise, dumping strategies can only outperform our synchronous strategy when $s < 2$, at which point the best dumping strategy has $\xi = m - 2$ and thus has larger $\nu$ when $$sn - \frac{n^2}{m} < s(m - 2) - \frac{(m - 2)^2}{m - 2} = s(m - 2) - (m - 2),$$ or $s(n + 2 - m) < \frac{n^2}{m} - m + 2$, which since $n + 2 - m > 0$ is equivalent to $s < \frac{n^2 - m^2 + 2m}{m(n + 2 - m)}$.

\begin{myremark}
    Note that we have not proven that these dumping strategies are optimal across all quantum strategies for any $p > 0$, only that they outperform synchronous strategies near the specified values of $s$ (or $p$).
    However, we also do not know of any strategy that outperforms both our synchronous strategy and all dumping strategies for any values of $n$, $m$, or $p$.
\end{myremark}

\subsection{Quantum Advantage over Classical Strategies}~\label{sub:quantumadvantage}
We can now apply the previous results to explicitly find the advantage provided by use of entanglement for this game.

Recall that the \textit{quantum advantage} is the difference between the values of the optimal quantum and classical strategies, and the \textit{quantum synchronous advantage} is the difference between the values of the optimal quantum and classical synchronous strategies. Also recall that we defined $r = n \bmod m$ and $q = \left\lfloor\frac{n}{m}\right\rfloor$, so that $n = qm + r$ and $r \in [0, m)$. 
\begin{mytheorem}~\label{thm:quantumsynchronousadvantage}
    The quantum synchronous advantage in this game is $$(1 - p)\lr{\frac{r(m - r)}{mn(n - 1)}}.$$
\end{mytheorem}
\begin{proof}
From Theorems \ref{thm:ourclassicalstrategyvalue} and \ref{thm:quantumsynchronousvalue}, we have that the classical and quantum synchronous values of this game are 
$$p + (1 - p)\lr{1 - \frac{(q - 1)n + (q + 1)r}{n(n - 1)}}$$
and
$$p + (1 - p)\lr{1 - \frac{n - m}{m(n - 1)}}$$
respectively. Thus, the quantum synchronous advantage is
\begin{align*}
&p + (1 - p)\lr{1 - \frac{n - m}{m(n - 1)}}- \lr{p + (1 - p)\lr{1 - \frac{(q - 1)n + (q + 1)r}{n(n - 1)}}}\\
& = (1 - p)\lr{\frac{(q - 1)n + (q + 1)r}{n(n - 1)} - \frac{n - m}{m(n - 1)}} \\
& = (1 - p)\lr{\frac{mqn - mn + mqr + mr - n^2 + nm}{mn(n - 1)}} \\
& = (1 - p)\lr{\frac{mqn + mqr + mr - n^2}{mn(n - 1)}} \\
& = (1 - p)\lr{\frac{mqn + (mq + r)r - r^2 + mr - n^2}{mn(n - 1)}} \\
& = (1 - p)\lr{\frac{mqn + nr - r^2 + mr - n^2}{mn(n - 1)}} \\
& = (1 - p)\lr{\frac{(mq + r)n - r^2 + mr - n^2}{mn(n - 1)}} \\
& = (1 - p)\lr{\frac{n^2 - r^2 + mr - n^2}{mn(n - 1)}} \\
& = (1 - p)\lr{\frac{r(m - r)}{mn(n - 1)}}
\end{align*}

\end{proof}

\begin{mytheorem}~\label{thm:quantumadvantage}
    When $p \geq \frac{2}{m+2}$, the quantum advantage in this game is $$(1 - p)\lr{\frac{r(m - r)}{mn(n - 1)}}.$$
\end{mytheorem}

\begin{proof}
When $p \geq \frac{2}{m+2}$, from Theorems \ref{thm:classicalvalue} and \ref{thm:quantumgamevalue}, we have that the classical and quantum synchronous values of this game are the same as the classical and quantum values, so by Theorem \ref{thm:quantumsynchronousadvantage} the quantum advantage is $$(1 - p)\lr{\frac{r(m - r)}{mn(n - 1)}}.$$
\end{proof}

\begin{myremark}
Note that this advantage is $0$ if and only if $r=0$ or, in other words, when $m \mid n$.
\end{myremark}

\begin{myremark}
$r(m - r) \le \frac{m}{2} \cdot \frac{m}{2} = \frac{m^2}{4}$, so the quantum advantage when $p \geq \frac{2}{m+2}$ satisfies $$(1 - p)\lr{\frac{r(m - r)}{mn(n - 1)}} \le (1 - p)\lr{\frac{m^2}{4mn(n - 1)}} = (1 - p)\lr{\frac{m}{4n(n - 1)}} = O\lr{\frac{m}{n^2}}.$$
\end{myremark}

\section{Conclusion}

The game we considered can be best thought of as the graph-coloring game on $K_n$ with $m$ colors, where $2 \le m < n$.
We found an optimal synchronous classical strategy 
and proved its optimality across all classical strategies when $p$, the probability that inputs are equal, is at least $\frac{2}{m+2}$. We then constructed an optimal synchronous quantum strategy
and proved its optimality across all quantum strategies when $p \geq \frac{2n-m}{(m+2)n-2m}$. We thus found the classical and quantum synchronous values of this game, and the general classical and quantum values for the aforementioned thresholds on $p$, as well as the quantum advantage over classical strategies (by difference in value) when $p \geq \frac{2}{m+2}$, which equals $0$ if and only if $m$ divides $n$.

\subsection{Open Problems}
We have not yet found which quantum strategies are optimal for $p < \frac{2n-m}{(m+2)n-2m}$, nor have we found a closed form for the classical optimum when $p < \frac{2}{m + 2}$.
We know that both classical and quantum synchronous strategies are suboptimal in at least some of these cases
but have found neither a general optimal strategy for these cases nor the value of $p$ below which synchronous strategies become suboptimal.

Another interesting question is if the optimal quantum success probability is achievable using less entanglement than our strategy (which uses $\log_2(m)$ ebits), or if $\log_2(m)$ is optimal. 

Finally, numerical simulations suggest that it may be possible to achieve the optimal quantum success probability with the same amount of entanglement using only real orthogonal transformations (as opposed to complex unitaries).

\subsection{Acknowledgements}
We thank Dr. Scott Aaronson (UT Austin) for suggesting we investigate this problem and providing continued support in revising this paper.

We also thank Dr. Venkat Arun (UT Austin) for discussing this problem with us and correcting an error in our initial construction.

\subsection{AI Use Statement}
ChatGPT and Claude were consulted at various points during this project to assist with ideation and literature review. In addition, Lemmas \ref{lem:operatorpsd}, \ref{lem:operatorinequality}, and \ref{lem:weakerxibound}, as well as Theorems \ref{thm:xisquaredinequality}, and \ref{thm:weakersynchoptimalatponehalf} were originally shown by a mixture of ChatGPT Astra 6 and Claude Opus 5.5 in response to our prompting. The final versions of all proofs in this paper are human-written and the authors take full responsibility for all work in this paper. 

\pagebreak

\begin{appendices}
\section{Deriving our Strategy}
  
We consider a particular class of strategies where Alice and Bob share the maximally entangled $m$-dimensional Bell state $\ket{\psi} := \frac{1}{\sqrt{m}}\sum_{k = 0}^{m - 1} \ket{k}\ket{k}$, apply the unitaries $A(x)$ and $B(y)$ when they receive inputs $x$ and $y$, respectively, then output the result of measuring their part in the computational basis. 

The state they jointly measure is then given by 
\begin{align*}\ket{\psi}_{xy}
& := (A(x) \otimes B(y))\ket{\psi} \\
& = \frac{1}{\sqrt{m}} \sum_{k = 0}^{m - 1} (A(x)\ket{k}) \otimes (B(y)\ket{k}) \\
& = \frac{1}{\sqrt{m}} \sum_{k = 0}^{m - 1} \lr{\lr{\sum_{a = 0}^{m - 1}A(x)_{ak}\ket{a}} \otimes \lr{\sum_{b = 0}^{m - 1}B(y)_{bk}\ket{b}}} \\
& = \frac{1}{\sqrt{m}} \sum_{k = 0}^{m - 1} \sum_{a = 0}^{m - 1}\sum_{b = 0}^{m - 1}A(x)_{ak}B(y)_{bk}\ket{a}\ket{b} \\
& = \frac{1}{\sqrt{m}} \sum_{a = 0}^{m - 1}\sum_{b = 0}^{m - 1}\sum_{k = 0}^{m - 1} A(x)_{ak}B(y)_{bk}\ket{ab} \\
& = \frac{1}{\sqrt{m}} \sum_{a = 0}^{m - 1}\sum_{b = 0}^{m - 1} \braket{A(x)^*_a \mid B(y)_b}\ket{ab} \\
& = \sum_{a = 0}^{m - 1}\sum_{b = 0}^{m - 1} \frac{1}{\sqrt{m}} \braket{A(x)^*_a \mid B(y)_b}\ket{ab}.
\end{align*} 

The probability that Alice and Bob give equal outputs is the sum of the squared amplitudes corresponding to basis states of the form $\ket{kk}$, so $$\xi = \sum_{x = 0}^{n - 1}\sum_{k = 0}^{m - 1} \lrb{\frac{1}{\sqrt{m}}\braket{A(x)^*_k \mid B(x)_k}}^2 = \frac{1}{m}\sum_{x = 0}^{n - 1}\sum_{k = 0}^{m - 1} \lrb{\braket{A(x)^*_k \mid B(x)_k}}^2$$ and $$\mu = \sum_{x = 0}^{n - 1}\sum_{y = 0}^{n - 1}\sum_{k = 0}^{m - 1} \lrb{\frac{1}{\sqrt{m}}\braket{A(x)^*_k \mid B(y)_k}}^2 = \frac{1}{m}\sum_{x = 0}^{n - 1}\sum_{y = 0}^{n - 1}\sum_{k = 0}^{m - 1} \lrb{\braket{A(x)^*_k \mid B(y)_k}}^2.$$

\subsection{Synchronous Quantum Strategies}
If $B(y) = A(x)^*$, the amplitude of $\ket{ab}$ is $\frac{1}{\sqrt{m}}\braket{B(y)_a\mid B(y)_b}$, which equals $\frac{1}{\sqrt{m}}$ if $a = b$ and $0$ otherwise, since $B(y)$ is unitary. Hence, Alice and Bob will always measure $\ket{kk}$ for some $k \in [m]$ and so give the same output. Therefore, any strategy with $B(x) = A(x)^*$ for all $x \in [n]$ is synchronous.

We then seek to minimize $\mu$, or equivalently its constant multiple $$m\mu = \sum_{x = 0}^{n - 1}\sum_{y = 0}^{n - 1}\sum_{k = 0}^{m - 1} \lrb{\braket{B(x)_k \mid B(y)_k}}^2.$$

\subsection{Vector Packing Problem}
Since $B(x)_k$ is a unit vector for each $x \in [n]$ and $k \in [m]$, we must have $$\sum_{x = 0}^{n - 1} \sum_{y = 0}^{n - 1} |\braket{B(x)_k \mid B(y)_k}|^2 > n$$ for all $k \in [m]$: otherwise, $\{B(x)_k \mid x \in [n]\}$ would be a set of $n$ orthogonal vectors in an $m$-dimensional space with $m < n$. 

As such, we attempt to minimize $\sum_{i = 0}^{n - 1} \sum_{j = 0}^{n - 1} |\braket{v_i \mid v_j}|^2$ across all sets of $n$ unit vectors $V \in (\mathbb{C}^m)^n \cong \C^{m \times n}$. This presents a ``vector packing'' problem, since our objective function discourages correlation between vectors.

By the definition of the matrix product, $\braket{v_i \mid v_j} = (V^\dagger V)_{ij}$ for all $i$ and $j$.
Tracing along the main diagonal (i.e., in the computational basis) gives $$\tr(V^\dagger V) = \sum_{i = 0}^{n - 1} (V^\dagger V)_{ii} = \sum_{i = 0}^{n - 1} \braket{v_i \mid v_i} = \sum_{i = 0}^{n - 1} ||v_i||^2  = \sum_{i = 0}^{n - 1} 1 = n.$$

By singular value decomposition, $V = U \Sigma W$ for some unitary $U \in \C^{m \times m}$, unitary $W \in \C^{n \times n}$, and nonnegative real diagonal $\Sigma \in \C^{m \times n}$, so $$V^\dagger V = (U \Sigma W)^\dagger(U \Sigma W) = W^\dagger\Sigma^\dagger U^\dagger U \Sigma W = W^\dagger\Sigma^T\Sigma W.$$

Hence, since traces are invariant under conjugation by unitaries like $W$, 
$$n = \tr(V^\dagger V) = \tr(W^\dagger \Sigma^T\Sigma W) = \tr(\Sigma^T\Sigma) = \sum_{i = 0}^{m - 1} (\Sigma_{ii})^2.$$

For all $i$ and $j$, $$
    |\braket{v_i \mid v_j}|^2 
    = \braket{v_i \mid v_j}\braket{v_i \mid v_j}^* 
    = \braket{v_i \mid v_j}\braket{v_j \mid v_i} 
    = (V^\dagger V)_{ij}(V^\dagger V)_{ji},
$$ so $$
\sum_{i = 0}^{n - 1} \sum_{j = 0}^{n - 1} |\braket{v_i \mid v_j}|^2
= \sum_{i = 0}^{n - 1} \sum_{j = 0}^{n - 1} (V^\dagger V)_{ij}(V^\dagger V)_{ji}
= \sum_{i = 0}^{n - 1} ((V^\dagger V)^2)_{ii}
= \tr((V^\dagger V)^2).
$$
Since $W$ is unitary, $WW^\dagger = \mathbb{I}$, so $$(V^\dagger V)^2 
= (W^\dagger\Sigma^T\Sigma W)^2 
= W^\dagger\Sigma^T\Sigma WW^\dagger\Sigma^T\Sigma W 
= W^\dagger(\Sigma^T\Sigma)^2W.$$
Thus, again using that the trace is invariant under conjugation by $W$,
$$\tr((V^\dagger V)^2) = \tr(W^\dagger(\Sigma^T\Sigma)^2W) = \tr((\Sigma^T\Sigma)^2) = \sum_{i = 0}^{m - 1} ((\Sigma^T\Sigma)^2)_{ii}.$$

$\Sigma$ and hence $\Sigma^T\Sigma$ and $(\Sigma^T\Sigma)^2$ are diagonal, so for all $i$, $$((\Sigma^T\Sigma)^2)_{ii} = ((\Sigma^T\Sigma)_{ii})^2 = ((\Sigma_{ii})^2)^2.$$

Letting $\lambda_i := \Sigma_{ii}^2$ for each $i \in [m]$, we then wish to minimize $\sum_{i = 0}^{m - 1} \lambda_i^2$ subject to the constraint $\sum_{i = 0}^{m - 1} \lambda_i = n$ (and $\lambda_i \ge 0$ for all $i \in [m]$).

By the Cauchy-Schwarz inequality on the standard inner product on $\R^m$, $$m\lr{\sum_{i = 0}^{m - 1} \lambda_i^2} 
 = \lr{\sum_{i = 0}^{m - 1} 1^2}\lr{\sum_{i = 0}^{m - 1} \lambda_i^2} \ge \lr{\sum_{i = 0}^{m - 1} 1 \cdot \lambda_i}^2 = \lr{\sum_{i = 0}^{m - 1} \lambda_i}^2 = n^2.$$

Hence, $\sum_{i = 0}^{m - 1} \lambda_i^2 \ge \frac{n^2}{m}$, and the inequality is saturated if and only if the vector with each entry $i$ equal to $\lambda_i$ is parallel to the vector with all entries equal to $1$, which requires that all $\lambda_i$ be equal to the same $\lambda \in \R$. Then using our constraint, $n = \sum_{i = 0}^{m - 1} \lambda_i = \sum_{i = 0}^{m - 1} \lambda = m\lambda$, so $\lambda = \frac{n}{m}$. Thus, $\Sigma_{ii} = \sqrt{\lambda_i} = \sqrt{\lambda} = \sqrt{\frac{n}{m}}$ for all $i \in [m]$.

However, we must choose $W$ to make every diagonal entry of $V^\dagger V$ equal $1$, not merely their sum equal $n$: i.e., $1 = (V^\dagger V)_{ii}
= (W^\dagger \Sigma^T\Sigma W)_{ii}$ for all $i$.

Since $\Sigma^T\Sigma$ is diagonal with $(\Sigma^T\Sigma)_{jj} = \Sigma_{jj}^2 = \lambda_j = \frac{n}{m}$ for $j < m$ and $(\Sigma^T\Sigma)_{jj} = 0$ for $j \ge m$,
$$(W^\dagger \Sigma^T\Sigma W)_{ii} 
= \sum_{j = 0}^{m - 1} (W^\dagger)_{ij} (\Sigma^T\Sigma)_{jj} W_{ji}
= \sum_{j = 0}^{m - 1} W_{ji}^* \lr{\frac{n}{m}} W_{ji}
= \frac{n}{m} \sum_{j = 0}^{m - 1} W_{ji}^*W_{ji}.$$ 
This sum is just the squared length of the vector given by the top $m$ entries of column $i$ of $W$. As such, our desired unitaries $W$ are exactly those whose uppermost $m \times n$ submatrices have column vectors of squared length $\frac{m}{n}$ each, which since the rows of a unitary must all have squared norm $1$ is the only possible length they could have if they all have equal length.

The matrix $U$ in the singular value decomposition of $V$ is unconstrained here, which makes sense because applying unitary operators to the columns of $V$ preserves their inner products and hence the value of the objective function. 
One obvious choice of $W$ that satisfies these constraints is the discrete Fourier transform matrix $F_n \in \C^{n \times n}$, where $(F_n)_{xy} := \frac{\omega_n^{xy}}{\sqrt{n}}$ with $\omega_n := e^{2\pi i/n}$: all its entries and thus all its truncated columns have equal magnitude.

\subsection{Trying the Fourier vectors}
Indeed, if $v_i := \sqrt{\frac{n}{m}}F_i$ is the normalized $i^\text{th}$ column of $F_n$ truncated to the top $m$ entries, 
\begin{align*} 
& \sum_{i = 0}^{n - 1}\sum_{j = 0}^{n - 1} |\braket{v_i \mid v_j}|^2 \\
& = \sum_{i = 0}^{n - 1}\sum_{j = 0}^{n - 1} \braket{v_i \mid v_j}^*\braket{v_i \mid v_j} \\
& =\sum_{i = 0}^{n - 1}\sum_{j = 0}^{n - 1} \lr{\frac{1}{m}\sum_{k = 0}^{m - 1}(\omega_n^{ki})^*\omega_n^{kj}}^*\lr{\frac{1}{m}\sum_{l = 0}^{m - 1} (\omega_n^{li})^*\omega_n^{lj}} \\
& = \frac{1}{m^2}\sum_{i = 0}^{n - 1}\sum_{j = 0}^{n - 1} \lr{\sum_{k = 0}^{m - 1} \omega_n^{ki - kj}}\lr{\sum_{l = 0}^{m - 1} \omega_n^{-li + lj}} \\
& = \frac{1}{m^2}\sum_{i = 0}^{n - 1}\sum_{j = 0}^{n - 1} \sum_{k = 0}^{m - 1} \sum_{l = 0}^{m - 1} \omega_n^{ki - kj}\omega_n^{-li + lj} \\
& = \frac{1}{m^2}\sum_{i = 0}^{n - 1} \sum_{k = 0}^{m - 1} \sum_{l = 0}^{m - 1} \sum_{j = 0}^{n - 1} \omega_n^{(j - i)(l - k)}.\end{align*}

If $l = k$, we have $\sum_{j = 0}^{n - 1} \omega_n^{(j - i)(l - k)} = \sum_{j = 0}^{n - 1} \omega_n^{0} = \sum_{j = 0}^{n - 1} 1 = n$. 

Otherwise, $\omega_n^{l - k} \neq 1$, so the terms in the sum interfere and
\begin{align*}
\sum_{j = 0}^{n - 1} \omega_n^{(j - i)(l - k)}
& = \sum_{j = 0}^{n - 1} \omega_n^{j(l -k)}\omega_n^{-i(l - k)}  \\
& = \omega_n^{-i(l - k)}\sum_{j = 0}^{n - 1} (\omega_n^{l - k})^{j}  \\
& = \omega_n^{-i(l - k)}\lr{\frac{1 - (\omega_n^{l - k})^n}{1 - \omega_n^{l - k}}} \\
& = \frac{\omega_n^{-i(l - k)}}{1 - \omega_n^{l - k}}\lr{1 - (\omega_n^n)^{l - k}} \\
& = \frac{\omega_n^{-i(l - k)}}{1 - \omega_n^{l - k}}\lr{1 - 1^{l - k}} = 0.
\end{align*}

Hence, $\sum_{j = 0}^{n - 1} \sum_{l = 0}^{m - 1} \omega_n^{(j - i)(l - k)} = n$ for all $i \in [n]$ and $k \in [m]$, so
$$\sum_{i = 0}^{n - 1} \sum_{k = 0}^{m - 1} \sum_{l = 0}^{m - 1} \sum_{j = 0}^{n - 1} \omega_n^{(j - i)(l - k)} = \sum_{i = 0}^{n - 1} \sum_{k = 0}^{m - 1} n = mn^2$$ and therefore $\sum_{i = 0}^{n - 1}\sum_{j = 0}^{n - 1} |\braket{v_i \mid v_j}|^2 = \frac{mn^2}{m^2} = \frac{n^2}{m}$, as desired.

\subsection{Extending to Unitary Bases}
To extend these vectors to a set of bases that minimize the objective, we let each $v_i$ be the first row of the basis $B(i)$. 
Since unitaries are closed under the operation of multiplying columns by phases, we hope that our bases will be identical apart from the Fourier phases applied to each column: then 
\begin{align*}
    \braket{B(x)_k \mid B(y)_k} 
    = \sum_{l = 0}^{m - 1} B(x)_{kl}^*B(y)_{kl} 
    = \sum_{l = 0}^{m - 1} (B(0)_{kl}\omega_n^{lx})^*B(0)_{kl}\omega_n^{ly}  
    = \sum_{l = 0}^{m - 1} |B(0)_{kl}|^2\omega_n^{l(y-x)}
\end{align*} for any row index $k$ and any inputs $x$ and $y$: the inner products depend only on the \textit{magnitudes} of the entries of the chosen ``template basis'' $B(0)$.

Hence, if each entry $B(0)_{kl}$ satisfies $|B(0)_{kl}| = |B(0)_{0l}| = \frac{1}{\sqrt{m}}$, 
$$\sum_{i = 0}^{n - 1} \sum_{j = 0}^{n - 1} |\braket{B(i)_k \mid B(j)_k}|^2 = \sum_{i = 0}^{n - 1} \sum_{j = 0}^{n - 1} |\braket{B(i)_0 \mid B(j)_0}|^2 = \frac{n^2}{m}$$
for each row index $k$, so our objective 
$\sum_{i = 0}^{n - 1} \sum_{j = 0}^{n - 1} \sum_{k = 0}^{m - 1} |\braket{B(i)_k \mid B(j)_k}|^2$ 
will equal
$$\sum_{k = 0}^{m - 1}\sum_{i = 0}^{n - 1} \sum_{j = 0}^{n - 1}  |\braket{B(i)_k \mid B(j)_k}|^2 = \sum_{k = 0}^{m - 1} \frac{n^2}{m} = m \cdot \frac{n^2}{m} = n^2.$$

We thus seek any $m \times m$ unitary whose every entry has magnitude $\frac{1}{\sqrt{m}}$.
A convenient such choice for $B(0)$ is the Fourier matrix $F_m \in \mathbb{C}^{m \times m}$.

\section{Proof of Synchronous Optimality for $p \geq \frac{1}{2}$}~\label{app:quantumponehalf} \begin{mylemma}~\label{lem:weakerxibound}
Any valid quantum strategy has $n\xi - \mu \leq n^2(1-\frac{1}{m})$.
\end{mylemma}
\begin{proof}
Let $\ket\psi$ be the state shared by Alice and Bob, and let $\{A_{x}^{a}\}_{a},\{B_{y}^{b}\}_{b}$ be Alice's and Bob's projective measurements, respectively. We then define the following symbols: $$u_{xa} = (A^a_x \otimes I) \ket\psi$$ $$v_{yb} = (I \otimes B^b_y ) \ket\psi$$ $$\overline{u}_a = \frac{1}{n}\sum_x u_{xa}$$ $$\overline{v}_b = \frac{1}{n}\sum_y v_{yb}$$
Looking at the above symbols, we can see that the probability that outputs are equal for a given pair of inputs $(x,y)$ is 
$$c_{xy} = \sum_a \Pr(a,a \mid x,y) = \sum_a \bra{\psi} A_x^a \otimes B_y^a \ket{\psi}$$
$A_x^a \otimes B_y^a = (A_x^a \otimes I) (I \otimes B_y^a)$ so 
$$\sum_a \bra{\psi} A_x^a \otimes B_y^a \ket{\psi} = \sum_a \bra{\psi} (A_x^a \otimes I) (I \otimes B_y^a) \ket{\psi}$$ 
By the adjoint property, this is equivalent to 
$$\sum_a \braket{ ((A_x^a)^\dagger \otimes I^\dagger)\psi \mid  (I \otimes B_y^a) \psi}$$
and since $A_x^a$ is a projector and thus Hermitian, $(A_x^a)^\dagger = (A_x^a)$, so we have 
$$\sum_a \braket{ ((A_x^a)^\dagger \otimes I^\dagger)\psi \mid  (I \otimes B_y^a) \psi} = \sum_a \braket{ (A_x^a \otimes I)\psi \mid  (I \otimes B_y^a) \psi} = \sum_a \braket{u_{xa} \mid v_{ya}} $$and thus
$$c_{xy} = \sum_a \braket{u_{xa} \mid v_{ya}}$$
It follows that 
$$\xi = \sum_{x,a} \braket{u_{xa} \mid v_{xa}}$$
and
$$\mu = \sum_{x,y,a} \braket{u_{xa} \mid v_{ya}}$$
Note that $\sum_y v_{ya} = n \overline{v}_a$, and thus $\sum_y \braket{u_{xa}\mid v_{ya}} = n\braket{u_{xa} \mid \overline{v}_a}$. Thus, 
$$\mu = \sum_{x,y,a} \braket{u_{xa} \mid v_{ya}} = n\sum_{x,a} \braket {u_{xa} \mid \overline{v}_a}$$

We can now find an expression for $n\xi - \mu$:
$$n\xi - \mu = n \sum_{x,a} \braket{u_{xa}\mid v_{xa}} - n\sum_{x,a} \braket{u_{xa}\mid\overline{v}_a} = n\sum_{x,a} \braket{u_{xa} \mid v_{xa}- \overline{v}_a}$$
Note that, for all $a$, $\sum_x \braket{\overline{u}_a \mid v_{xa}-\overline{v}_a} = \braket{\overline{u}_a\mid n\overline{v}_a-n\overline{v}_a} = \braket{\overline{u}_a \mid 0} = 0$, which means that we can use it as a zero quantity and freely subtract it, so
$$n\xi - \mu = n\sum_{x,a} \braket{u_{xa}\mid v_{xa}- \overline{v}_a} = n\sum_{x,a} \braket{u_{xa}\mid v_{xa}- \overline{v}_a} - n\sum_{x,a} \braket{\overline{u}_a \mid v_{xa}-\overline{v}_a}$$
Thus
$$ n\xi - \mu = n\sum_{x,a} \braket{u_{xa} - \overline{u}_a\mid v_{xa}- \overline{v}_a}$$
Note that $\xi, \mu$ are both sums of probabilities, and thus must be entirely real. As such, $\mathrm{Re}(n\sum_{x,a} \braket{u_{xa} - \overline{u}_a\mid v_{xa}- \overline{v}_a}) = n\sum_{x,a} \braket{u_{xa} - \overline{u}_a\mid v_{xa}- \overline{v}_a}$, so 
$$ n\xi - \mu = n\sum_{x,a} \mathrm{Re}\braket{u_{xa} - \overline{u}_a\mid v_{xa}- \overline{v}_a}$$

We will return to that expression in a little while. In the meantime, recall that $A_x^a$ is a projector, and thus $||u_{xa}||^2 = \bra{\psi}(A_x^a)^2 \otimes I \ket{\psi} = \bra{\psi}(A_x^a) \otimes I \ket{\psi}$. This tells us that, by completeness of projective measurement, $$\sum_a||u_{xa}||^2 = 1$$and thus $$\sum_{x,a} ||u_{xa}||^2 = n$$

We will use that soon. But first, we will establish a lower bound on $\sum_a||\overline{u}_a||^2$. By the generalized triangle inequality, $$||\sum_{a}\overline{u}_a|| \leq \sum_a||\overline{u}_a||$$Since both sides are nonnegative real numbers, this tells us that $$||\sum_{a}\overline{u}_a||^2 \leq (\sum_a||\overline{u}_a||)^2$$

By the Cauchy-Schwarz inequality, since there are $m$ values of $a$, $$m \sum_a||\overline{u}_a||^2 \geq (\sum_a||\overline{u}_a||)^2 \geq ||\sum_{a}\overline{u}_a||^2$$ and thus 

$$\sum_a||\overline{u}_a||^2 \geq \frac{||\sum_{a}\overline{u}_a||^2}{m}$$

We have
$$|| \sum_{a} \overline{u}_a|| = || \sum_{x, a}  \frac{u_{xa}}{n}|| = ||\frac{1}{n}\sum_{x} (\sum_{a} u_{xa})|| = ||\frac{1}{n}\sum_{x} \ket{\psi}|| = ||\ket{\psi}|| = 1,$$ so

$$\sum_a||\overline{u}_a||^2 \geq \frac{1^2}{m} = \frac{1}{m}$$

Again, that will be used soon. Next, we define 
$$D_u = \sum_{x,a} ||u_{xa} - \overline{u}_a||^2$$
We can expand the term inside the sum to get us
$$D_u = \sum_{x,a} (||u_{xa}||^2 - 2\mathrm{Re}\braket{u_{xa}\mid \overline{u}_a} + ||\overline{u}_a||^2)$$
Distributing the sum over $x$ gives us $$D_u = \sum_{a} (\sum_x(||u_{xa}||^2) - 2\sum_x(\mathrm{Re}\braket{u_{xa}\mid \overline{u}_a}) + \sum_x(||\overline{u}_a||^2))$$
$$= \sum_{a} (\sum_x(||u_{xa}||^2) - 2\sum_x(\mathrm{Re}\braket{u_{xa}\mid \overline{u}_a}) + n||\overline{u}_a||^2)$$
$$=\sum_{a} (\sum_x(||u_{xa}||^2) - 2(\mathrm{Re}\sum_x(\braket{u_{xa}\mid \overline{u}_a})) + n||\overline{u}_a||^2)$$
$$=\sum_{a} (\sum_x(||u_{xa}||^2) - 2(\mathrm{Re}(n\braket{\overline{u}_{a}\mid \overline{u}_a})) + n||\overline{u}_a||^2)$$
$$=\sum_{a} (\sum_x(||u_{xa}||^2) - 2n||\overline{u}_{a}||^2 + n||\overline{u}_a||^2)$$
$$=\sum_{a} (\sum_x(||u_{xa}||^2) - n||\overline{u}_{a}||^2)$$
And we can distribute the sum over $a$ to find 
$$D_u = \sum_{x,a}(||u_{xa}||^2) - n\sum_a||\overline{u}_{a}||^2$$We established above that $\sum_{x,a}(||u_{xa}||^2) = n$ so 
$$D_u = n - n\sum_a||\overline{u}_{a}||^2$$ and we also established above that $\sum_a||\overline{u}_a||^2 \geq \frac{1}{m}$, which tells us that $$D_u = n - n\sum_a||\overline{u}_{a}||^2 \leq n - \frac{n}{m} = n(1-\frac{1}{m})$$
Importantly, note that none of our steps here used any property specific to Alice, so the same inequality holds true for Bob's state and projectors: 
$$D_v = \sum_{y,b} ||v_{yb} - \overline{v}_b||^2 \leq n(1-\frac{1}{m})$$

At last, we can come back to our expression for $n\xi - \mu$:
$$n\xi - \mu= n\sum_{x,a} \mathrm{Re}\braket{u_{xa} - \overline{u}_a\mid v_{xa}- \overline{v}_a}$$
As a simple property of absolute values, 
$$n\sum_{x,a} \mathrm{Re}\braket{u_{xa} - \overline{u}_a\mid v_{xa}- \overline{v}_a} \leq n|\sum_{x,a} \mathrm{Re}\braket{u_{xa} - \overline{u}_a\mid v_{xa}- \overline{v}_a}|$$
By the triangle inequality,
$$|\sum_{x,a} \mathrm{Re}\braket{u_{xa} - \overline{u}_a\mid v_{xa}- \overline{v}_a}| \leq \sum_{x,a} |\mathrm{Re}\braket{u_{xa} - \overline{u}_a\mid v_{xa}- \overline{v}_a}|$$
The magnitude of a complex value is guaranteed to be at least the magnitude of its real part, so 
$$\sum_{x,a} |\mathrm{Re}\braket{u_{xa} - \overline{u}_a\mid v_{xa}- \overline{v}_a}| \leq \sum_{x,a} |\braket{u_{xa} - \overline{u}_a\mid v_{xa}- \overline{v}_a}|$$
For each pair $(x,a)$ we can apply Cauchy-Schwarz to see that $$|\braket{u_{xa} - \overline{u}_a\mid v_{xa}- \overline{v}_a}| \leq ||u_{xa} - \overline{u}_a|| \cdot ||v_{xa}- \overline{v}_a||$$ so 
$$\sum_{x,a} |\braket{u_{xa} - \overline{u}_a\mid v_{xa}- \overline{v}_a}| \leq \sum_{x,a} ||u_{xa} - \overline{u}_a|| \cdot ||v_{xa}- \overline{v}_a||$$
Once more using Cauchy-Schwarz we can observe that 
$$(\sum_{x,a} ||u_{xa} - \overline{u}_a|| \cdot ||v_{xa}- \overline{v}_a||) ^2 \leq \sum_{x,a}(||u_{xa} - \overline{u}_a||)^2  \sum_{x,a}(||v_{xa}- \overline{v}_a||)^2 = D_uD_v$$ 
and thus 
$$\sum_{x,a} ||u_{xa} - \overline{u}_a|| \cdot ||v_{xa}- \overline{v}_a||  \leq \sqrt{D_u}\sqrt{D_v}$$
This allows us to conclude that 
$$n\xi - \mu= n\sum_{x,a} \mathrm{Re}\braket{u_{xa} - \overline{u}_a\mid v_{xa}- \overline{v}_a} \leq n \sqrt{D_u}\sqrt{D_v}$$and since we know that $D_u \leq n(1-\frac{1}{m})$ and $D_v \leq n(1-\frac{1}{m})$, this tells us that 
$$n\xi - \mu \leq n^2(1 - \frac{1}{m})$$
\end{proof}

\begin{mytheorem}~\label{lem:weakerquantumvaluebound}
At $p = \frac{1}{2}$, any quantum strategy has $\omega \le \frac{1}{2} + \frac{1}{2}\lr{1-\frac{(n-m)}{m(n-1)}}$.
\end{mytheorem}

\begin{proof}

Recall that the value for a given strategy can be written as 
$$\omega = 1 - p + \frac{1 - p}{n(n - 1)}\nu$$
where 
$$\nu := \lr{\frac{p(n - 1)}{1 - p} + 1}\xi - \mu$$

Since $p = \frac{1}{2}$, we have 
$$ \frac{p(n-1)}{1-p}+1 = \frac{\frac{1}{2}(n-1)}{\frac{1}{2}} + 1 = n$$
so $\nu = n \xi - \mu$ and thus by Lemma \ref{lem:weakerxibound}
$\nu \leq n^2(1-\frac{1}{m})$. Thus 
$$\omega = 1 - p + \frac{1 - p}{n(n - 1)}\nu \leq 1 - p + \frac{1 - p}{n(n - 1)}(n^2)(1-\frac{1}{m})$$
Substituting in $p = \frac{1}{2}$, we have
$$\omega \leq 1 - \frac{1}{2} + \frac{\frac{1}{2}}{n(n - 1)}(n^2)(1-\frac{1}{m}) = \frac{1}{2} + \frac{1}{2}(\frac{n-\frac{n}{m}}{n-1})$$

We have $$\frac{n-\frac{n}{m}}{n-1} = \frac{mn-n}{m(n-1)} = \frac{mn-m + m - n}{m(n-1)} = 1 - \frac{n-m}{m(n-1)}.$$
Substituting into the above inequality for $\omega$ gives our desired bound.
\end{proof}

\begin{mytheorem}~\label{thm:weakersynchoptimalatponehalf}
Our quantum synchronous strategy is optimal at $p = \frac{1}{2}$ (i.e., when equal inputs are exactly as likely as unequal inputs).
\end{mytheorem}
\begin{proof}
By Theorem \ref{thm:quantumstrategyvalue}, our quantum synchronous strategy's value is $$p + (1-p)(1-\frac{n-m}{m(n-1)})$$
Substituting in $p = \frac{1}{2}$ this gives us 
$$\frac{1}{2} + \frac{1}{2}(1-\frac{n-m}{m(n-1)})$$
By Theorem \ref{lem:weakerquantumvaluebound}, for $p = \frac{1}{2}$, we have $$\omega \leq \frac{1}{2} + \frac{1}{2}(1-\frac{(n-m)}{m(n-1)})$$
for all quantum strategies, and thus our synchronous strategy achieves the upper bound on win probability for $p = \frac{1}{2}$.
\end{proof}

\begin{mytheorem}\label{thm:weakersynchstrategyoptimalmorethanonehalf}
Our quantum synchronous strategy is optimal for all $p \in [\frac{1}{2}, 1]$.
\end{mytheorem}
\begin{proof}
This follows directly from Theorems \ref{thm:synchoptimalatpmeansoptimalabovep} and \ref{thm:weakersynchoptimalatponehalf}: our strategy is optimal over all quantum strategies for $p=\frac{1}{2}$ by Theorem \ref{thm:weakersynchoptimalatponehalf}, and thus by Theorem \ref{thm:synchoptimalatpmeansoptimalabovep} our strategy is optimal for all $p \in [\frac{1}{2},1]$.
\end{proof}

\begin{mytheorem}~\label{thm:weakeroptimalquantumvalueforpatleastonehalf}
The optimal quantum value for this game, for $p \in [\frac{1}{2}, 1]$, is $p + (1-p)(1-\frac{n-m}{m(n-1)})$.
\end{mytheorem}
\begin{proof}
This follows directly from Theorem \ref{thm:weakersynchstrategyoptimalmorethanonehalf}: since our strategy is optimal over all quantum strategies for $p \in [\frac{1}{2}, 1]$, and since we have by Theorem \ref{thm:quantumstrategyvalue}  the success probability of our strategy is $$p + (1-p)(1-\frac{n-m}{m(n-1)})$$ that must be the optimal quantum value over all quantum strategies for $p \in [\frac{1}{2}, 1]$.
\end{proof} 

\end{appendices}

\bibliographystyle{plain}
\bibliography{main}

\end{document}